\input harvmac
\input epsf

\font\myrm=cmr10 scaled 600
\font\mymi=cmmi10 scaled 600
\font\mysevenrm=cmr7 scaled 600
\font\myfiverm=cmr5 scaled 600
\font\myfivemi=cmmi5 scaled 600
\font\mysevenmi=cmmi7 scaled 600

\newcount\figno
\figno=0 
\def\fig#1#2#3{
\par\begingroup\parindent=0pt\leftskip=1cm\rightskip=1cm\parindent=0pt
\baselineskip=11pt
\global\advance\figno by 1
\midinsert
\epsfxsize=#3
\centerline{\epsfbox{#2}}
\vskip 12pt
{\bf Fig.\ \the\figno: } #1\par
\endinsert\endgroup\par
}
\def\figlabel#1{\xdef#1{\the\figno}}

\def\tb{{\tilde{b}}}

\def\tX{{\widetilde{X}}}

\def\p{\partial}
\def\IP{\relax{\rm I\kern-.18em P}}
\def\IR{\relax{\rm I\kern-.18em R}}
\def\IK{\relax{\rm I\kern-.18em K}}

\lref\seven{
N.~R.~Constable, D.~Z.~Freedman, M.~Headrick, S.~Minwalla, L.~Motl,
A.~Postnikov and W.~Skiba,
``PP-wave string interactions from perturbative Yang-Mills theory,''
JHEP {\bf 0207}, 017 (2002)
[arXiv:hep-th/0205089].
}

\lref\LeeCU{
P.~Lee and J.~w.~Park,
``Open strings in PP-wave background from defect conformal field theory,''
arXiv:hep-th/0203257.
}

\lref\BerkovitsZV{
N.~Berkovits,
``Conformal field theory for the superstring in a Ramond-Ramond plane
wave background,''
JHEP {\bf 0204}, 037 (2002)
[arXiv:hep-th/0203248].
}

\lref\MetsaevRE{
R.~R.~Metsaev and A.~A.~Tseytlin,
``Exactly solvable model of superstring in plane wave Ramond-Ramond
background,''
Phys.\ Rev.\ D {\bf 65}, 126004 (2002)
[arXiv:hep-th/0202109].
}

\lref\BMN{
D.~Berenstein, J.~M.~Maldacena and H.~Nastase,
``Strings in flat space and pp waves from N = 4 super Yang Mills,''
JHEP {\bf 0204}, 013 (2002)
[arXiv:hep-th/0202021].
}

\lref\MetsaevBJ{
R.~R.~Metsaev,
``Type IIB Green-Schwarz superstring in plane wave Ramond-Ramond background,''
Nucl.\ Phys.\ B {\bf 625}, 70 (2002)
[arXiv:hep-th/0112044].
}

\lref\GreenXX{
M.~B.~Green and J.~H.~Schwarz,
``Supersymmetrical Dual String Theory. 2. Vertices And Trees,''
Nucl.\ Phys.\ B {\bf 198}, 252 (1982).
}

\lref\GreenFU{
M.~B.~Green and J.~H.~Schwarz,
``Superstring Field Theory,''
Nucl.\ Phys.\ B {\bf 243}, 475 (1984).
}

\lref\GreenTK{
M.~B.~Green and J.~H.~Schwarz,
``Extended Supergravity In Ten-Dimensions,''
Phys.\ Lett.\ B {\bf 122}, 143 (1983).
}

\lref\GreenTC{
M.~B.~Green and J.~H.~Schwarz,
``Superstring Interactions,''
Nucl.\ Phys.\ B {\bf 218}, 43 (1983).
}

\lref\GSW{
M.~B.~Green, J.~H.~Schwarz and E.~Witten,
``Superstring Theory,'' Cambridge University Press, Chapter 11.}

\lref\GSB{
M.~B.~Green, J.~H.~Schwarz and L.~Brink,
``Superfield Theory Of Type II Superstrings,''
Nucl.\ Phys.\ B {\bf 219}, 437 (1983).
}

\lref\MandelstamHK{
S.~Mandelstam,
``Interacting String Picture Of The Neveu-Schwarz-Ramond Model,''
Nucl.\ Phys.\ B {\bf 69}, 77 (1974).
}

\lref\MandelstamWH{
S.~Mandelstam,
``Interacting String Picture Of The Fermionic String,''
Prog.\ Theor.\ Phys.\ Suppl.\  {\bf 86}, 163 (1986).
}

\lref\GursoyTX{
U.~Gursoy, C.~Nunez and M.~Schvellinger,
``RG flows from Spin(7), CY 4-fold and HK manifolds to AdS, Penrose
limits and pp waves,''
JHEP {\bf 0206}, 015 (2002)
[arXiv:hep-th/0203124].
}

\lref\TakayanagiHV{
T.~Takayanagi and S.~Terashima,
``Strings on orbifolded pp-waves,''
JHEP {\bf 0206}, 036 (2002)
[arXiv:hep-th/0203093].
}

\lref\CveticHI{
M.~Cvetic, H.~Lu and C.~N.~Pope,
``Penrose limits, pp-waves and deformed M2-branes,''
arXiv:hep-th/0203082.
}

\lref\BlauNE{
M.~Blau, J.~Figueroa-O'Farrill, C.~Hull and G.~Papadopoulos,
``A new maximally supersymmetric background of IIB superstring theory,''
JHEP {\bf 0201}, 047 (2002)
[arXiv:hep-th/0110242].
}

\lref\MaldacenaRE{
J.~M.~Maldacena,
``The large N limit of superconformal field theories and supergravity,''
Adv.\ Theor.\ Math.\ Phys.\  {\bf 2}, 231 (1998)
[Int.\ J.\ Theor.\ Phys.\  {\bf 38}, 1113 (1999)]
[arXiv:hep-th/9711200].
}

\lref\HatsudaKX{
M.~Hatsuda, K.~Kamimura and M.~Sakaguchi,
``Super-PP-wave algebra from super-AdS $\times S$ algebras
in eleven-dimensions,''
Nucl.\ Phys.\ B {\bf 637}, 168 (2002)
[arXiv:hep-th/0204002].
}

\lref\DasCW{
S.~R.~Das, C.~Gomez and S.~J.~Rey,
``Penrose limit, spontaneous symmetry breaking and holography in pp-wave
background,''
Phys.\ Rev.\ D {\bf 66}, 046002 (2002)
[arXiv:hep-th/0203164].
}

\lref\FloratosUH{
E.~Floratos and A.~Kehagias,
``Penrose limits of orbifolds and orientifolds,''
JHEP {\bf 0207}, 031 (2002)
[arXiv:hep-th/0203134].
}

\lref\KimFP{
N.~w.~Kim, A.~Pankiewicz, S.~J.~Rey and S.~Theisen,
``Superstring on pp-wave orbifold from large-N quiver gauge theory,''
Eur.\ Phys.\ J.\ C {\bf 25}, 327 (2002)
[arXiv:hep-th/0203080].
}

\lref\AlishahihaEV{
M.~Alishahiha and M.~M.~Sheikh-Jabbari,
``The PP-wave limits of orbifolded $AdS_5 \times S^5$,''
Phys.\ Lett.\ B {\bf 535}, 328 (2002)
[arXiv:hep-th/0203018].
}

\lref\ItzhakiKH{
N.~Itzhaki, I.~R.~Klebanov and S.~Mukhi,
``PP wave limit and enhanced supersymmetry in gauge theories,''
JHEP {\bf 0203}, 048 (2002)
[arXiv:hep-th/0202153].
}

\lref\BlauMW{
M.~Blau, J.~Figueroa-O'Farrill and G.~Papadopoulos,
``Penrose limits, supergravity and brane dynamics,''
Class.\ Quant.\ Grav.\  {\bf 19}, 4753 (2002)
[arXiv:hep-th/0202111].
}

\lref\BerensteinZW{
D.~Berenstein, E.~Gava, J.~M.~Maldacena, K.~S.~Narain and H.~Nastase,
``Open strings on plane waves and their Yang-Mills duals,''
arXiv:hep-th/0203249.
}

\lref\PandoZayasRX{
L.~A.~Pando Zayas and J.~Sonnenschein,
``On Penrose limits and gauge theories,''
JHEP {\bf 0205}, 010 (2002)
[arXiv:hep-th/0202186].
}

\lref\RussoRQ{
J.~G.~Russo and A.~A.~Tseytlin,
``On solvable models of type IIB superstring in NS-NS and R-R plane wave
backgrounds,''
JHEP {\bf 0204}, 021 (2002)
[arXiv:hep-th/0202179].
}

\lref\GomisKM{
J.~Gomis and H.~Ooguri,
``Penrose limit of N = 1 gauge theories,''
Nucl.\ Phys.\ B {\bf 635}, 106 (2002)
[arXiv:hep-th/0202157].
}

\lref\BlauDY{
M.~Blau, J.~Figueroa-O'Farrill, C.~Hull and G.~Papadopoulos,
``Penrose limits and maximal supersymmetry,''
Class.\ Quant.\ Grav.\  {\bf 19}, L87 (2002)
[arXiv:hep-th/0201081].
}

\lref\GubserTV{
S.~S.~Gubser, I.~R.~Klebanov and A.~M.~Polyakov,
``A semi-classical limit of the gauge/string correspondence,''
Nucl.\ Phys.\ B {\bf 636}, 99 (2002)
[arXiv:hep-th/0204051].
}

\lref\DabholkarZC{
A.~Dabholkar and S.~Parvizi,
``Dp branes in pp-wave background,''
Nucl.\ Phys.\ B {\bf 641}, 223 (2002)
[arXiv:hep-th/0203231].
}

\lref\CveticSI{
M.~Cvetic, H.~Lu and C.~N.~Pope,
``M-theory pp-waves, Penrose limits and supernumerary supersymmetries,''
Nucl.\ Phys.\ B {\bf 644}, 65 (2002)
[arXiv:hep-th/0203229].
}

\lref\ChuIN{
C.~S.~Chu and P.~M.~Ho,
``Noncommutative D-brane and open string in pp-wave background with  B-field,''
Nucl.\ Phys.\ B {\bf 636}, 141 (2002)
[arXiv:hep-th/0203186].
}

\lref\MichelsonWA{
J.~Michelson,
``(Twisted) toroidal compactification of pp-waves,''
Phys.\ Rev.\ D {\bf 66}, 066002 (2002)
[arXiv:hep-th/0203140].
}

\lref\BilloFF{
M.~Billo and I.~Pesando,
``Boundary states for GS superstrings in an $Hpp$ wave background,''
Phys.\ Lett.\ B {\bf 536}, 121 (2002)
[arXiv:hep-th/0203028].
}

\lref\SkenderisVF{
K.~Skenderis and M.~Taylor,
``Branes in AdS and pp-wave spacetimes,''
JHEP {\bf 0206}, 025 (2002)
[arXiv:hep-th/0204054].
}

\lref\BakRQ{
D.~s.~Bak,
``Supersymmetric branes in PP wave background,''
arXiv:hep-th/0204033.
}

\lref\KumarPS{
A.~Kumar, R.~R.~Nayak and Sanjay,
``D-brane solutions in pp-wave background,''
Phys.\ Lett.\ B {\bf 541}, 183 (2002)
[arXiv:hep-th/0204025].
}

\lref\LuKW{
H.~Lu and J.~F.~Vazquez-Poritz,
``Penrose limits of non-standard brane intersections,''
Class.\ Quant.\ Grav.\  {\bf 19}, 4059 (2002)
[arXiv:hep-th/0204001].
}

\lref\GauntlettCS{
J.~P.~Gauntlett and C.~M.~Hull,
``pp-waves in 11-dimensions with extra supersymmetry,''
JHEP {\bf 0206}, 013 (2002)
[arXiv:hep-th/0203255].
}

\lref\HatsudaXP{
M.~Hatsuda, K.~Kamimura and M.~Sakaguchi,
``From super-AdS${}_5 \times $S${}_5$ algebra to super-pp-wave algebra,''
Nucl.\ Phys.\ B {\bf 632}, 114 (2002)
[arXiv:hep-th/0202190].
}

\lref\LeighPT{
R.~G.~Leigh, K.~Okuyama and M.~Rozali,
``PP-waves and holography,''
Phys.\ Rev.\ D {\bf 66}, 046004 (2002)
[arXiv:hep-th/0204026].
}

\lref\KiritsisKZ{
E.~Kiritsis and B.~Pioline,
``Strings in homogeneous gravitational waves and null holography,''
JHEP {\bf 0208}, 048 (2002)
[arXiv:hep-th/0204004].
}

\Title{\vbox{\baselineskip12pt
        \hbox{hep-th/0204146}
        \hbox{PUPT-2028}
        \hbox{HUTP-02/A011} 
}}{Superstring Interactions in a pp-wave Background}

\centerline{
Marcus Spradlin${}^{1}$ and Anastasia Volovich${}^{2}$
}

\bigskip
\centerline{${}^{1}$~Department of Physics}
\centerline{Princeton University}
\centerline{Princeton, NJ 08544}
\centerline{\tt spradlin@feynman.princeton.edu}
\centerline{}
\centerline{${}^{2}$~Department of Physics}
\centerline{Harvard University}
\centerline{Cambridge, MA 02138}
\centerline{\tt nastya@gauss.harvard.edu}

\vskip .3in
\centerline{\bf Abstract}

We construct light-cone gauge superstring field theory in a pp-wave
background with Ramond-Ramond flux.
The leading term in the interaction Hamiltonian
is determined up
to an overall function of $p^+$ by requiring closure of the
pp-wave superalgebra.  The bosonic and fermionic Neumann matrices
for this cubic vertex
are derived, as is the interaction point operator.
We comment on the development of a $1/\mu p^+$ expansion for these
results.

\smallskip

\Date{}

\listtoc
\writetoc

\newsec{Introduction}

Recently Berenstein, Maldacena and Nastase \BMN\ have argued that
a sector of ${\cal{N}}=4$ SU(N) Yang-Mills theory
containing operators
with large R-charge $J$
is dual to IIB superstring theory on a certain
gravitational plane wave background (pp-wave) with Ramond-Ramond flux.
The pp-wave solution of type IIB supergravity is
\BlauNE
\eqn\ppsol{
ds^2 = -4 dx^+ dx^- - \mu^2 x_I x^I (dx^+)^2 + dx_I dx^I, \qquad
F_{+1234} = F_{+5678} \propto \mu,
}
where $I=1,\ldots,8.$
It has 32 supersymmetries and can be obtained as
a Penrose limit from $AdS_5 \times S^5$ \refs{\BlauDY}.
This application of the AdS/CFT correspondence \MaldacenaRE\ 
is particularly exciting
because the string worldsheet theory in this background
is exactly solvable, as shown by Metsaev and Tseytlin
\refs{\MetsaevBJ, \MetsaevRE},
whereas in the more familiar
$AdS \times S$  dualities it is difficult to go beyond
the supergravity approximation on the string theory side.
Furthermore, in this case there are two expansion parameters 
$\lambda'={g^2_{YM} N \over J^2}={1 \over (\mu p^+ \alpha')^2}$
and $g_2={J^2 \over N}=4 \pi g_s (\mu p^+ \alpha')^2$,
so there exists the intriguing possibility that there is a regime
in which both sides of the `duality' are perturbative \seven.

The authors of \BMN\ obtained the light-cone gauge
worldsheet action for a single free
string in the pp-wave background by 
summing certain planar diagrams.
It is natural to ask whether
string interactions can be incorporated into this picture.
In this paper we study the light-cone
vertex
corresponding to the
joining of two closed strings into one closed string (see Figure~1)
in the pp-wave background.
Our analysis follows closely the corresponding flat space calculation
of Green, Schwarz and Brink \GSB\ (see also \GSW).
We construct the light-cone gauge superstring field theory for this
background and derive expressions for the Neumann matrices and 
interaction point operator.
Note that our calculation is purely on the `string side,' although
it is our hope to eventually compare with results from  the
`field theory side.'

\fig{Strings 1 and 2 merge to form string 3 via a local interaction
    at the point marked X on the worldsheet.  Note that
    light-cone time increases from right to left
in our diagrams.}{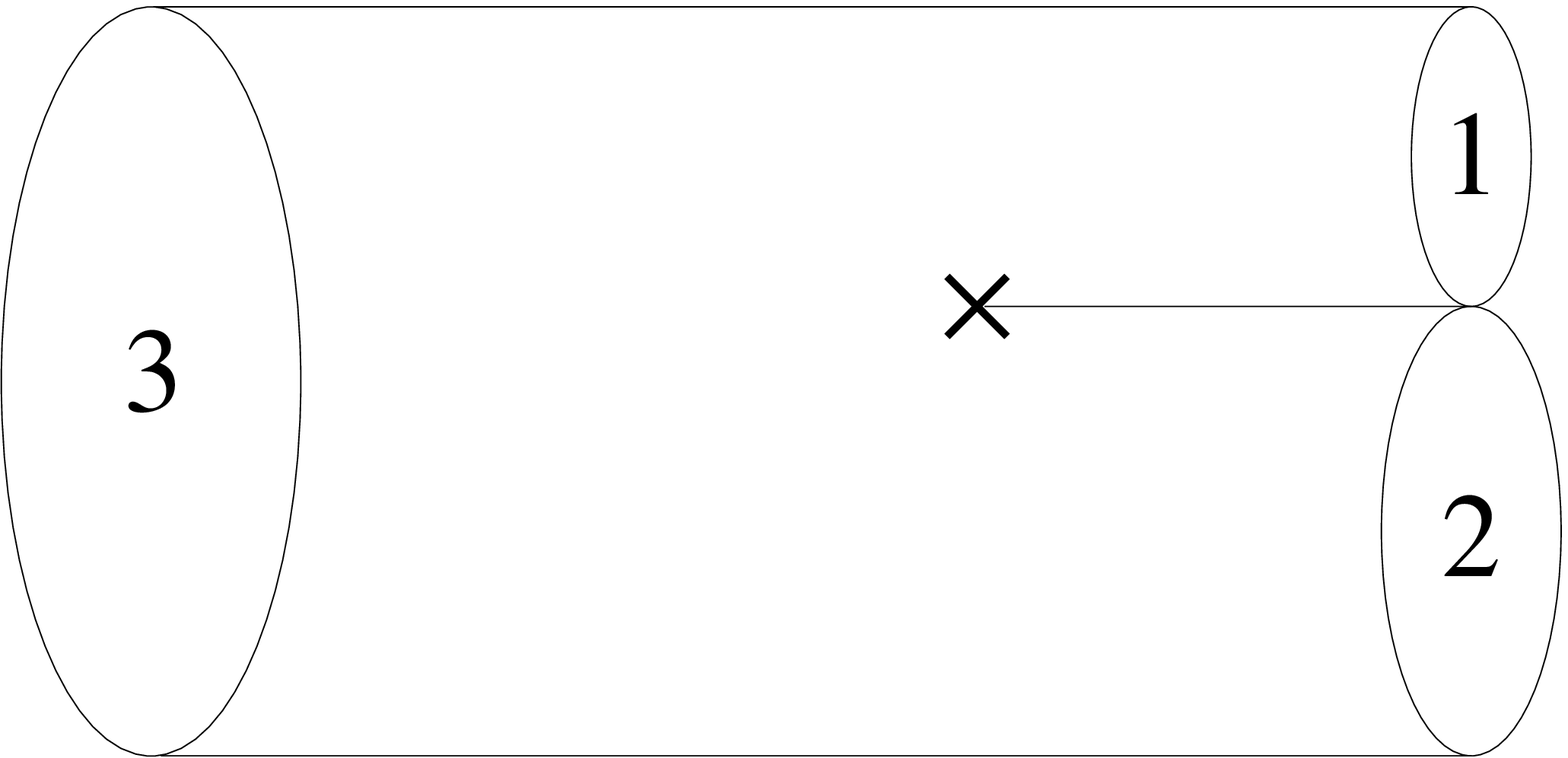}{1.5in}

Light-cone string field theory is an old subject, harking back
to the work of Mandelstam \refs{\MandelstamHK,\MandelstamWH}
and extensively developed for the superstring
by Green and Schwarz \refs{\GreenTK
\GreenTC \GreenXX-\GreenFU}, and
for the closed type IIB string by Green, Schwarz
and Brink \GSB.  The light-cone formulation is especially appropriate
for the pp-wave background because of the beautiful way in which
light-cone `string bits' naturally emerge in the field
theory description \BMN\ (see Figure~2).
Although there are various equivalent approaches to studying
string interactions, it is clear that any attempt to connect
field theory results to string theory results benefits from
consideration of light-cone gauge.

\fig{A typical field theory diagram which corresponds to the
string diagram of Figure~1 and represents the merging 
of $\Tr[Z^{J_2} \phi^2] \Tr[Z^{J_1} \phi^2]$
into
$\Tr[Z^{J_3} \phi^4]$.
(We have been very schematic---see \BMN\ for a detailed description
of the identification between field theory operators and string states.)
}{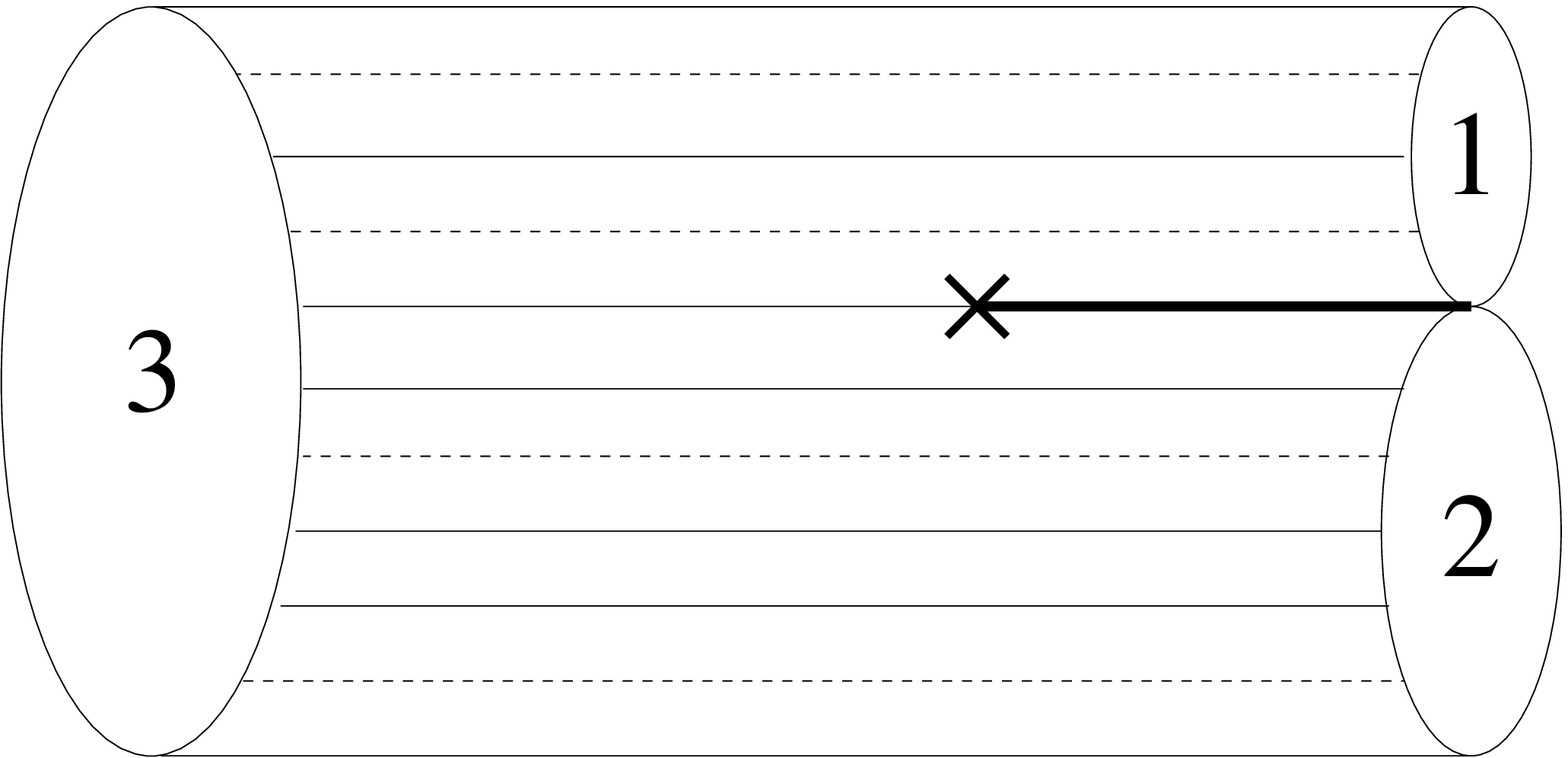}{1.5in}

Furthermore, a significant advantage of the light-cone approach
is that it is particularly clean (at least conceptually)
in the pp-wave background.\foot{Modulo the difficulties associated
with states having $p^+=0$,
which are a perennial nuisance of light-cone theories.}
In a general background,
it is unknown even how to determine what the observables of string
theory are.
In flat space we have the S-matrix and
in AdS we have boundary correlation functions.  
In the pp-wave background, all string modes are massive in light-cone gauge,
so a general state is labelled by $p^+$ and an infinite number of
harmonic oscillator occupation numbers.
In string perturbation theory, the observables of this theory
are simply quantum mechanical transition amplitudes.
This paper reports the first non-trivial term in
the interaction Hamiltonian for this theory.

Having extolled the virtues of light-cone gauge, let us now be
honest about some of its limitations.
In flat space, the cubic string vertex is determined uniquely by
the requirement that the nonlinearly realized super-Poincar\'e
algebra closes to first order in the string coupling.
The pp-wave superalgebra  has the same number of supersymmetries
as the flat space algebra,
but it has fewer bosonic generators.  In
particular, there are no symmetries analogous to the $J^{+-}$ or $J^{-I}$
symmetries of flat space.
Thus there are no symmetries which relate states of different
values of $p^+$, so the pp-wave superalgebra is sufficient only
to determine the cubic vertex up to a
function $f(p^+_{(1)}, p^+_{(2)},
p^+_{(3)})$ of the values of $p^+$ for the three strings.
(Of course $p^+$ is still conserved, so $\sum p^+_{(r)} = 0$.)
Since this factor is independent of the particular string states
in question, one could derive this factor once and for all from
some other method---for example, by comparing with a supergravity
calculation (which would be valid for $\mu p^+ \alpha' \ll 1$)
or with the dual field theory (for $\mu p^+ \alpha' \gg 1)$.
Some other technical difficulties of working in light-cone gauge
have been highlighted by Berkovits \BerkovitsZV, who has constructed
a quantizable, covariant,
conformal worldsheet action for the superstring in a
pp-wave background.

The plan of the paper is the following.
In section 2 we review free string theory in the pp-wave background.
In section 3 we explain how the interaction vertex is
determined in light-cone string field theory \`a la Green,
Schwarz and Brink.
Then in section 4 we get our feet wet by studying the interaction
just of the supergravity modes.
In section 5 we derive the bosonic and fermionic Neumann matrices which
couple all of the string modes, and in section 6 we construct the
operators which supersymmetry requires to be inserted at
the interaction point.  Finally in section 7 we discuss
the problem of developing a $1/\mu p^+$ expansion for comparison
with calculations on the field theory side.

A number of recent papers have considered related
issues, including obtaining pp-wave string theory from field theories
with ${\cal{N}}=1$ supersymmetry \refs{\ItzhakiKH
\GomisKM -
\PandoZayasRX}, orbifolded pp-waves \refs{\AlishahihaEV
\KimFP
\TakayanagiHV -
\FloratosUH}, holography in the pp-wave background
\refs{
\DasCW
\KiritsisKZ-
\LeighPT
},
and other aspects of Penrose limits and strings
in plane wave backgrounds
\refs{
\BlauMW
\RussoRQ  
\CveticHI
\GursoyTX
\MichelsonWA
\CveticSI
\GauntlettCS -
\LuKW
\GubserTV
}.

\newsec{Free String Theory}

In this section we review 
the Hilbert space and Hamiltonian for
a free string on the pp-wave background \ppsol.  This has been
discussed in \refs{\MetsaevBJ,\MetsaevRE}, but it is useful to include
a review here in order to fix notation and especially to display
explicitly the nontrivial basis transformation required to write
the fermionic part of the Hamiltonian in a canonical form.

We consider a string 
with fixed light-cone momentum $p^+ \ne 0$.
We allow for $p^+$ to be negative, so that in the following
sections when we consider string interactions, strings with $p^+ > 0$
($p^+ < 0$) will be in incoming (outgoing) strings.
Following \GSB, we  set $\alpha' = 2$ and use the notation
$\alpha = 2 p^+$, $e(\alpha) = {\rm sign}(\alpha)$.

\subsec{Bosonic Sector}

The light-cone action for the bosons is
\eqn\aaa{
S = {e(\alpha) \over 8 \pi} \int d\tau \int_0^{2 \pi |\alpha|} d\sigma\ 
[ \p_+ x^I \p_- x^J - \mu^2 x^I x^J ] \delta_{IJ},
}
where $\p_\pm = \p_\tau \pm \p_\sigma$ and $I=1,\ldots,8$ labels
the directions transverse to the light-cone.
Since the bosonic sector of the theory is SO(8) invariant, we will
frequently omit the $I$ index in order to avoid unnecessary clutter.

Let us expand the string coordinate $x(\sigma)$ and momentum density
$p(\sigma)$ in Fourier modes $x_n$ and $p_n$ according to the formula
\eqn\xandp{
\eqalign{
x(\sigma) &= x_0 + {1 \over \sqrt{2}} \sum_{n \ne 0}
 ( x_{|n|} - i e(n) x_{-|n|})
e^{i n \sigma/|\alpha|},\cr
p(\sigma) &= {1 \over 2 \pi |\alpha|}
\left[ p_0 + 
{1 \over \sqrt{2}}
\sum_{n \ne 0} (p_{|n|} - i e(n) p_{-|n|}) e^{i n \sigma/|\alpha|}
\right],
}}
where $n$ ranges from $-\infty$ to $+\infty$.
The complicated form of the coefficients has been chosen so that
the modes
$x_n$ and $p_n$ are associated to Hermitian operators
$\hat{x}_n$ and $\hat{p}_n$.
We will use capital letters $X(\sigma)$, $P(\sigma)$ to denote
the operators given by
\xandp\ with $x_n$ and $p_n$ replaced
by $\hat{x}_n$ and $\hat{p}_n$.
The normalization is chosen so that the canonical
commutation relation
\eqn\aaa{
[X(\sigma), P(\sigma')] = i \delta(\sigma-\sigma')
}
follows from imposing
\eqn\aaa{
[\hat{x}_m, \hat{p}_n] = i \delta_{mn}.
}
The coefficients $p_n$ in the expansion \xandp\ are related to the
coefficients used in \GSB\ by
\eqn\transcribe{
p_0 \leftrightarrow p, \qquad p_n \leftrightarrow \sqrt{n} p_n^{\rm I},
\qquad p_{-n} \leftrightarrow  \sqrt{n} p_n^{\rm II},
}
for $n>0$.  This relation will help us transcribe some of the
results of \GSB\ with little difficulty.
The basis \xandp\ is more convenient for the
pp-wave since there is no need
to treat the zero mode separately.
All of the string oscillations---left-movers, right-movers,
and the zero modes---can be treated on an equal footing.

The Hamiltonian is
\eqn\hghd{
h = {e(\alpha) \over 2} \int_0^{2 \pi |\alpha|} d\sigma \left[
4 \pi p^2 + {1 \over 4 \pi} ( (\p_\sigma x)^2 + \mu^2 x^2)\right].
}
Inserting $X(\sigma)$ and $P(\sigma)$  into \hghd\ gives
the operator version
\eqn\hamiltonian{
H = {1 \over \alpha} \sum_{n=-\infty}^\infty
\left[ \hat{p}_n^2 + {1 \over 4} \omega_n^2 \hat{x}_n^2 \right],
}
where $\omega_n=\sqrt{n^2+(\alpha \mu)^2}.$
We can write this in a canonical normal-ordered form
by introducing
the lowering operators\foot{We caution the reader that this
is not the familiar string theory basis employed by \BMN.
Their operators are related to ours by
$a^{\rm BMN}_n = {1 \over \sqrt{2}}( a_{|n|} - i e(n) a_{-|n|})$ for
$n\ne 0$.}
\eqn\ladder{
a_n = {1 \over \sqrt{\omega_n}} \hat{p}_n -
{i \over 2} \sqrt{\omega_n} \hat{x}_n}
which obey
\eqn\aaa{
[ a_m, a_n^\dagger] = \delta_{mn}.
}
Then the Hamiltonian may be written as
\eqn\hamilcanon{
H = {1 \over \alpha}\left[\sum_{n=-\infty}^\infty \omega_n a_n^\dagger
a_n + A\right],
}
where
\eqn\bnormal{
A = {\delta{}^I{}_I \over 2}
\sum_{n=-\infty}^\infty \omega_n
}
is a normal-ordering constant which will cancel against the fermionic
contribution in the next subsection.

\subsec{Fermionic Sector}

The action is
\eqn\aaa{
S = {1 \over 8 \pi} \int d\tau \int_0^{2 \pi |\alpha|} d\sigma\ [i(\bar{\theta}
\p_\tau \theta + \theta \p_\tau \bar{\theta}) - \theta \p_\sigma \theta
+ \bar{\theta} \p_\sigma \bar{\theta} - 2
\mu \bar{\theta} \Pi \theta].
}
Here $\theta^a$ is a complex positive chirality 
SO(8) spinor (although we will frequently omit the spinor index
$a=1,\ldots,8$) and $\Pi = \gamma^1 \gamma^2 \gamma^3 \gamma^4$.
We expand
$\theta$ and its conjugate momentum\foot{Actually,
$\lambda$ is $-i$ times the conjugate momentum to $\theta$.} as
$\lambda \equiv {1 \over 4 \pi}
\bar{\theta}$
as\foot{The coefficients are related to those of \GSB\ by
$ \lambda_0=\lambda, ~\lambda_n =\lambda_n^{\rm I},~
\lambda_{-n}=\lambda_n^{\rm II}$
for $n>0$.}
\eqn\tandl{\eqalign{
\theta(\sigma)&= \vartheta_0 + {1 \over \sqrt{2}}
\sum_{n \ne 0} ( \vartheta_{|n|} - i e(n) \vartheta_{-|n|})
e^{i n \sigma/|\alpha|},\cr
\lambda(\sigma) &= 
{1 \over 2 \pi |\alpha|} \left[ \lambda_0 + {1 \over \sqrt{2}}
\sum_{n \ne 0} (\lambda_{|n|} - i e(n) \lambda_{-|n|})
e^{i n \sigma/|\alpha|}\right],
}}
with the reality condition $\lambda_n^* = {|\alpha|
\over 2} \vartheta_n$.
The anticommutation relation
\eqn\aaa{
\{ \Theta^a(\sigma), \Lambda^b(\sigma') \} =
\delta^{ab}
\delta(\sigma - \sigma')
}
follows from
\eqn\aaa{
\{ \hat{\vartheta}_m^a,
\hat{\lambda}_n^b \} = \delta^{a b} \delta_{mn}.
}

The Hamiltonian is
\eqn\fham{
h = 
\ha \int_0^{2 \pi |\alpha|}
d\sigma\ \left[-4 \pi  \lambda \p_\sigma
\lambda + {1 \over 4 \pi} \theta \p_\sigma \theta + 2  \mu
(\lambda \Pi \theta)\right].
}
In terms of the mode operators,
we have
\eqn\sdfkj{
H=\ha \sum_{n=-\infty}^\infty\left[{n
\over 2}\left({4 \over \alpha^2} \hat{\lambda}_{-n}
\hat{\lambda}_n - \hat{\vartheta}_{-n}
\hat{\vartheta}_n\right)
+2  \mu \hat{\lambda}_n \Pi \hat{\vartheta}_n
\right].
}
We would like to write the Hamiltonian in canonical form in
terms of fermionic lowering
operators $b_n$, $n=-\infty,\ldots,\infty$, satisfying
\eqn\aaa{
\{ b_m, b_n^\dagger \} = \delta_{mn}
}
(with spinor indices $a,b$ suppressed).
Because of
the last term in \sdfkj, the required change of basis is nontrivial.
For $n>0$ we define
\eqn\iofshe{
\eqalign{
\hat{\vartheta}_{n}&= {1 \over \sqrt{|\alpha|}} c_n
\left[ (1+\rho_n \Pi) b_n + e(\alpha) (1-\rho_n \Pi)
b_{-n}^\dagger \right],\cr
\hat{\vartheta}_{-n}&= {1 \over \sqrt{|\alpha|}} c_n
\left[(1+\rho_n \Pi) b_{-n}-
e(\alpha) (1-\rho_n \Pi)
b_n^\dagger\right],\cr
\hat{\lambda}_{n}&= {\sqrt{|\alpha|} \over 2} c_n
\left[e(\alpha) (1-\rho_n \Pi) b_{-n} +
 (1+\rho_n \Pi) b_n^\dagger\right],\cr
\hat{\lambda}_{-n}&=  {\sqrt{|\alpha|} \over 2} c_n
\left[-e(\alpha) (1-\rho_n \Pi)b_n+ 
(1+\rho_n \Pi) b_{-n}^\dagger\right],
}}
where 
\eqn\aaa{
\rho_n = {\omega_n - n \over \alpha \mu}, \qquad c_n =
{1 \over \sqrt{1 + \rho_n^2}},
}
and for the zero-modes we take
\eqn\aaa{\eqalign{
\hat{\vartheta}_0 &= {1 \over \sqrt{2|\alpha|}}
\left[ (1 + e(\alpha) \Pi) b_0 +  (1 -
e(\alpha) \Pi) b_0^\dagger\right],\cr
\hat{\lambda}_0 &= \ha \sqrt{|\alpha| \over 2}
\left[(1 + e(\alpha)
\Pi) b_0^\dagger +  (1 - e(\alpha) \Pi) b_0\right].
}}
In terms of the $b_n$, the Hamiltonian takes the desired simple form
\eqn\aaa{
H=
{1 \over \alpha}
\left[
\sum_{n=-\infty}^\infty \omega_n b_n^\dagger b_n 
+ B\right],
}
where the normal-ordering constant
\eqn\fnormal{
B = - {\delta{}^a{}_a \over 2}
\sum_{n=-\infty}^\infty \omega_n
}
cancels \bnormal\ as advertised.

\subsec{The Hilbert Space}

We have expressed the normal-ordered free string Hamiltonian
\eqn\aaa{
H = {1 \over \alpha}\sum_{n=-\infty}^\infty \omega_n (\delta_{IJ}
a_n^{I \dagger} a_n^J + \delta_{ab}
b_n^{a\dagger} b_n^b),\qquad
\omega_n = \sqrt{n^2 + (\alpha \mu)^2}
}
in terms of an infinite number of canonically normalized raising
and lowering operators.
The vacuum is defined by
\eqn\vacdef{
a_n |{\rm vac}\rangle = b_n|{\rm vac}\rangle = 0, \qquad \forall n,
}
and a general state is constructed by acting on $|{\rm vac}\rangle$
with the creation operators $a_n^\dagger$ and $b_n^\dagger$.
The subspace ${\cal{H}}_1$ of physical states
is obtained by imposing the constraint
\eqn\physical{
\sum_{n=-\infty}^\infty n ( \delta_{IJ} a_{-n}^{I \dagger}
a_n^J + \delta_{ab}
b_{-n}^{a \dagger} b_n^b )|\Psi\rangle = 0,
}
which expresses the fact that there is no significance to the
choice of origin for the $\sigma$ coordinate on the string
worldsheet.

The state $|{\rm vac}\rangle$ defined
by \vacdef\ is not the ``ground state'' that we will be using
for the rest of this paper.  A more convenient choice
for $|0\rangle$
\MetsaevRE\ is a state which is the ground
state for all the bosonic oscillators as well as the non-zero
fermionic oscillators, but satisfies
\eqn\aaa{
\vartheta_0|0\rangle = 0.
}
The energy of this state is $H |0 \rangle = 4 \mu |0\rangle$, so the
zero should be thought of signifying the occupation number, and not
the energy.
The energy can be raised (for $\alpha>0$)
by applying to $|0\rangle$ any of the four
components of $(1 + \Pi)
\hat{\lambda}_0$, or lowered
by applying any of the four components of $(1 - \Pi) \hat{\lambda}_0$.
The Hilbert spaces built on $|0\rangle$ and $|{\rm vac}\rangle$ are of course
isomorphic and differ just by a relabeling of the states.

\subsec{Symmetries of the Free Theory}

The pp-wave superalgebra is a contraction of the $AdS_5 \times S^5$
superalgebra \refs{\HatsudaXP, \HatsudaKX}.
The isometries are
generated by $H$, $P^I$, $P^+$, $J^{+I}$, $J^{ij}$ and $J^{i'j'}$,
where the indices run over
$I=1,\ldots,8$, $i=1,\ldots,4$ and $i'=5,\ldots,8$.
There are also 32 supercharges $Q^+$, $\overline{Q}{}^+$, $Q^-$ and
$\overline{Q}{}^-$.
The interesting (anti)-commutation relations
(i.e., those which are not the same as in flat space)
are
\eqn\superone{
[H, P^I] = i\mu^2 J^{+I}, \qquad [ P^I, Q^-_{\dot{a}} ] = \mu
(\Pi \gamma^I)_{\dot{a} b} Q^+_b, \qquad
[H, Q^+_a] = \mu \Pi_{a b} Q^+_b,
}
\eqn\supertwo{
\{ Q^-_{\dot{a}}, \overline{Q}{}^-_{\dot{b}} \} = 2 \delta_{\dot{a} \dot{b}} H
+ i \mu (\gamma_{ij} \Pi)_{\dot{a} \dot{b}} J^{ij}
+ i \mu (\gamma_{i'j'} \Pi)_{\dot{a} \dot{b}} J^{i' j'},
}
with similarly complicated formulas for
$\{ Q^+, \overline{Q}{}^-\}$ and $\{Q^-, \overline{Q}{}^+\}$ (see for
example \MetsaevRE).
We do not write the rest of the algebra since we will have
no need for the precise form of these relations.
We have only written \superone\ and \supertwo\ to highlight
a very important fact about the pp-wave superalgebra, which is
that although the algebra satisfied by the free generators
is complicated, we will see that
the algebra satisfied by the interaction
terms in $H$, $Q^-$ and $\overline{Q}{}^-$ (when they
are promoted to operators which act non-linearly
on the full string theory Hilbert space)
satisfy essentially the same algebra as in flat space,
in a sense made precise in the following section.

We will however make
use of the supercharges
\eqn\qplus{
q^+ = \int_0^{2 \pi |\alpha|} d \sigma\ \sqrt{2}\,\lambda, \qquad
\overline{q}{}^+ = \int_0^{2 \pi |\alpha|}d\sigma\,
{e(\alpha) \over 2 \pi \sqrt{2}} \theta
}
and
\eqn\qminus{\eqalign{
q^- &=
\int_0^{2 \pi |\alpha|} d\sigma \ \left[{4 \pi e(\alpha)}
p^I \gamma_I \lambda
- {i \over 4 \pi} \p_\sigma x^I \gamma_I \theta
- i \mu x^I \gamma_I \Pi \lambda
\right],\cr
\overline{q}{}^- &= \int_0^{2 \pi |\alpha|} d\sigma
\left[ p^I \gamma_I \theta +
i{e(\alpha)}
\p_\sigma x^I \gamma_I  \lambda
+ {i \over 4 \pi} e(\alpha) \mu x^I \gamma_I \Pi \theta
\right].
}}

\newsec{Interacting String Theory}

The primary purpose of this section is to explain in detail
what it is that we are trying to calculate in this paper.
We review the light-cone string field theory
formulation used by Green, Schwarz and Brink \GSB\ to calculate
the cubic string interaction in flat space.
The only essential
conceptual
differences in the pp-wave case are that the superalgebra is smaller,
and that in this case
even the zero modes of the string have nonzero energy.\foot{The
term ``zero modes'' is nevertheless not a misnomer, since we mean
the modes which are the zeroth Fourier coefficient in the
$\sigma$ coordinate on the worldsheet.}

\subsec{Light-cone String Field Theory}

The $m$-string Hilbert space ${\cal{H}}_m$
is the product of $m$ copies of the single string Hilbert
space ${\cal{H}}_1$ described in the previous section.
The Hilbert space of the full string theory is the
sum ${\cal{H}} = |{\rm vacuum}\rangle \oplus
{\cal{H}}_1 \oplus {\cal{H}}_2 \oplus \cdots$.
The basic object in string field theory is the field operator
$\Phi$ which can create or destroy complete strings.
In a momentum space
representation, $\Phi$ is a function of $x^+$ (the light-cone time),
$\alpha$, and the worldsheet momentum
densities $p^I(\sigma)$ and
$\lambda^a(\sigma)$.
To be specific, we can expand $\Phi$ in the number-basis representation
as (we suppress the fermionic degrees of freedom and the transverse
index $I$ for simplicity)
\eqn\sfield{
\Phi[p(\sigma)]
= \sum_{ \{ n_k \} } \varphi_{\{ n_k \}} \prod_{k=-\infty}^\infty
\psi_{n_k}(p_k).
}
Here $p_k$ is the $k$-th Fourier mode of
$p(\sigma)$, as in the expansion \xandp.
The sum is over all possible sets of harmonic oscillator
occupation numbers $\{n_k\}$ which satisfy the physical state
condition \physical\ 
and $\psi_n(p)$ is the harmonic oscillator wavefunction for
occupation number $n$.
Finally $\varphi_{\{n_k\}}$ is an operator ${\cal{H}}_m \to {\cal{H}}_{m
\pm 1}$ that
creates  (if $\alpha<0$) or destroys (if $\alpha>0$) a string
in the state $|\{n_k\}\rangle$ at time $\tau = 0$.

The Hamiltonian as well as the other
generators of the pp-wave superalgebra from the previous
section
are promoted to operators on the full
Hilbert space by expressing them in terms of the string field $\Phi$.
For example, the free Hamiltonian is
\eqn\hfield{
H_2 = \ha \int \alpha d\alpha\, D^8p(\sigma) D^8\lambda(\sigma)\ \Phi^\dagger
h \Phi,
}
where $h$ is the sum of the bosonic and fermionic
contributions, \hghd\ and \fham.
The subscript $2$ signifies that this term is quadratic in
string fields.  
The leading interaction $H_3$ which we
seek to determine has an integrand which is cubic in $\Phi$.

\subsec{Symmetries of the Interaction}

There are two essentially different kinds of symmetries.
The kinematical generators
\eqn\kine{
P^+, P^I, J^{+I}, J^{ij}, J^{i'j'}, Q^+, \overline{Q}{}^+
}
do not involve $\p \over \p x^+$ and therefore act at fixed
light-cone time.  These generators are quadratic in the string
field, as in \hfield, and receive no corrections
in the interacting theory.
In other words, they act diagonally on the Hilbert space,
mapping ${\cal{H}}_{(r)} \to {\cal{H}}_{(r)}$, where
${\cal{H}}_{(r)}$ is the Hilbert space of the $r$-th string.
On the other hand, the dynamical generators
\eqn\dynam{
H, Q^-, \overline{Q}{}^-
}
are corrected by interactions, so that they can create
or annihilate strings.
The full Hamiltonian
has the form
\eqn\hamil{
H = H_2 + \kappa H_3 + \cdots,
}
with similar expansions for $Q^-$ and $\overline{Q}{}^-$,
where $\kappa$ is the  coupling constant.\foot{The question
of what $\kappa$ actually {\it is} in terms of other parameters, such
as $g_s$, $\mu$, $p^+$, $\alpha'$, $\pi$, or $2$,
is irrelevant to our analysis.
We simply use $\kappa$ as an auxiliary parameter to keep
track of the perturbative expansion.}
Our goal in this paper will be to construct the leading terms $H_3$,
$Q^-_3$ and $\overline{Q}{}^-_3$ by requiring that the
generators \kine\ and \dynam\ satisfy the pp-wave superalgebra
to order ${\cal{O}}(\kappa)$.
Presumably
one could fix the higher order terms
in a similar fashion, although we do not consider them in this paper.

Let us make a couple of general observations about
the cubic interaction.
First, we note that
although transverse momentum is not conserved in the pp-wave
background \ppsol\ because $\p/\p x^I$ is not a Killing vector,
the
${\cal{O}}(\kappa)$ terms of the first relation in
\superone\ imply that
\eqn\supertwo{
[ H_3, P^I ] = 0.
}
(Indeed the same relation holds for any $H_k$ with  $k>2$.)
Therefore,
the interaction Hamiltonian is translationally invariant---it is
only the free Hamiltonian $H_2$ which breaks this symmetry
(by confining particles in a harmonic oscillator potential).
The cubic interaction must therefore contain a $\Delta$-functional
which implements conservation of transverse momentum locally on the
worldsheet:
\eqn\aone{
\Delta^8 \left[ \sum_{r=1}^3 p^I_{(r)} (\sigma)\right],
}
where $r$ labels the three strings.
Similarly,
we see from \superone\ that
although $Q^+$ does not commute with the free Hamiltonian $H_2$,
it does commute with the interaction $H_3$, so the latter must
also contain the $\Delta$-functional
\eqn\atwo{
\Delta^8 \left[ \sum_{r=1}^3 \lambda^a_{(r)} (\sigma)\right].
}
Finally, $P^+$ is a good quantum number in the pp-wave background,
implying the delta-function
\eqn\athree{
\delta(\alpha_{(1)} + \alpha_{(2)} + \alpha_{(3)}).
}

To summarize, we have gotten a lot of mileage out of the fact that
although the pp-wave superalgebra \superone\ looks quite different from
the flat space super-Poincar\'e algebra, those differences only
affect the leading ($\kappa = 0$) terms in the dynamical generators.
The interaction terms all (anti)-commute with the kinematical
generators \kine.  Much of the machinery of \GSB\ therefore
carries over to the present case.

\subsec{Number Basis Vertex}

The conservation laws \aone, \atwo\ and \athree\ imply that the
cubic interaction can be written in the form
\eqn\hthreeint{
H_3 = \int d \mu_3\ h_3(\alpha_{(r)}, p_{(r)}(\sigma), x'_{(r)}(\sigma),
\lambda_{(r)}(\sigma)) \Phi(1) \Phi(2) \Phi(3),
}
where $\Phi(r) = \Phi[x^+, \alpha_{(r)}, p_{(r)}(\sigma),
\lambda_{(r)}(\sigma)]$ is the string
field for string $r$, $h_3$ is a factor to be determined, and
the measure is
\eqn\measure{
d\mu_3 = \left( \prod_{r=1}^3 d \alpha_{(r)} D^8 p_{(r)}(\sigma)
D^8 \lambda_{(r)}(\sigma) \right) \delta( {\textstyle{\sum}}
\alpha_{(r)}) \Delta^8 \left[ {\textstyle{\sum}} p_{(r)} (\sigma)\right]
\Delta^8 \left[ {\textstyle{\sum}} \lambda_{(r)} (\sigma)\right].
}
There will be similar expressions for $Q^-_3$ and $\overline{Q}{}^-_3$,
involving different factors $q_3$ and $\overline{q}_3$
but the same measure \measure.  Therefore it
is fair that an entire section of this paper
is devoted exclusively to the study of this measure, which
is determined just by consideration of the kinematical symmetries
\kine, while we postpone
discussion of the prefactors $h_3$, $q_3$ and $\overline{q}_3$,
which are determined by more complicated dynamical considerations, until
section 6.

The delta function of $\sum \alpha_{(r)}$ guarantees that
$H_3$ creates or annihilates at most a single string.
Therefore it is sufficient to consider the action of $H_3$
from
${\cal{H}}_1 \to {\cal{H}}_2$ (or the adjoint of this process),
since any other strings
simply go along for the ride.
It is convenient to express $H_3$
not as an operator ${\cal{H}}_1 \to {\cal{H}}_2$
but rather as a state $|H\rangle$ in ${\cal{H}}_3$ via the
identification\foot{In IIB string theory, the superfield $\Phi$
satisfies a generalized reality condition \GreenTK\ which implies that
the adjoint on ${\cal{H}}$ is not the same as the one
induced from the adjoint on ${\cal{H}}_1 \subset {\cal{H}}$.
The prime on string 3 here denotes the string field theory adjoint
of the single-string state $|3\rangle$.}
\eqn\state{
\langle 3 | H | 1 \rangle |2 \rangle
=  \langle 1 | \langle 2 | \langle 3'|H\rangle,
}
Similarly we identify $Q_3^-$ and $\overline{Q}{}_3^-$ with states
$|Q^-\rangle$, $|\overline{Q}{}^-\rangle$ in ${\cal{H}}_3$.

Using an identity which is roughly of the form
\eqn\aaa{
\sum_{n=0}^\infty |n\rangle \psi_n(p) \sim 
\exp \left( - {1 \over 4} p^2 + p a^\dagger - \ha a^\dagger a^\dagger
\right) |0\rangle \equiv \psi(p) |0\rangle
}
(this particular $\psi(p)$ is for an oscillator with $\omega=4$,
as an example)
to relate the momentum wavefunctions
$\psi_n(p)$
which appear in the string field expansion \sfield\ 
to number-basis states (and a similar
relation for fermionic oscillators), it is straightforward
\GreenTC\ to show that the
operator \hthreeint\ can be expressed as a state
in the three string Hilbert space
by the formula
(again we suppress the fermions and transverse indices)
\eqn\vertex{
|H\rangle = 
\left[
\int d\mu_3\ h_3 \prod_{r=1}^3 \prod_{k=-\infty}^\infty \psi(p_{k(r)})\right]
|0\rangle.
}
Since we postpone discussion of the prefactor $h_3$, we will actually
calculate first the three-string vertex
just with the kinematical delta-functions \measure,
\eqn\aaa{
|V\rangle \equiv \left[
\int d\mu_3 \prod_{r=1}^3 \prod_{k=-\infty}^\infty \psi(
p_{k(r)}) \right] |0\rangle.
}
These delta-functions are common to the three dynamical
generators, which can therefore be represented as
\eqn\aaa{
|H \rangle = \hat{h}_3 |V\rangle, 
\qquad
|Q^- \rangle = \hat{q}_3 |V \rangle, \qquad
|\overline{Q}{}^- \rangle = \hat{\overline{q}}_3 |V \rangle
}
for some operators $\hat{h}_3$, $\hat{q}_3$ and $\hat{\overline{q}}_3$
which will be determined in section 6.

\fig{
Parameterization of the $\sigma$ coordinate for the interaction
of Figure~1.  The interaction occurs at the points marked $I$,
which are identified.
}
{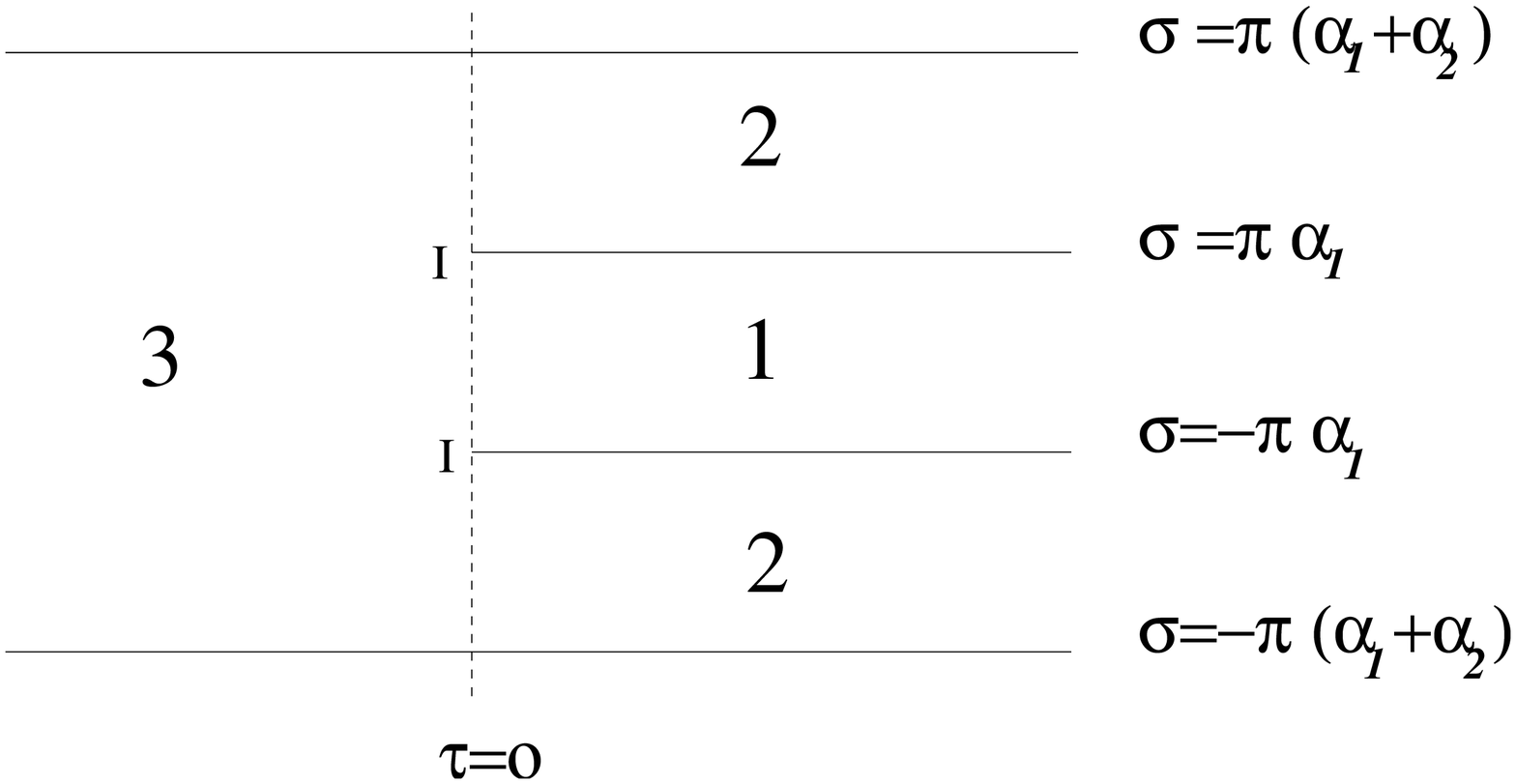}{2.2in}

\subsec{Parametrization of the Interaction}

We consider three strings joining as in Figure~2.  The strings
are labelled by $r=1,2,3$ and in light-cone gauge their widths are
$2 \pi|\alpha_{(r)}|$, with
$\alpha_{(1)} + \alpha_{(2)} + \alpha_{(3)} = 0$.
We will take $\alpha_{(1)}$ and $\alpha_{(2)}$ positive
for purposes of calculation,
although the final vertex will allow for all permutations of
the process depicted in Figure~1.
The coordinates of the three strings may be  parametrized by
\eqn\sigmas{\eqalign{
\sigma_{(1)}
&= \sigma \qquad\qquad~~~ -\pi \alpha_{(1)} \le \sigma \le \pi \alpha_{(1)},\cr
\sigma_{(2)} &= \left\{\matrix{\sigma - \pi \alpha_{(1)} & 
\pi \alpha_{(1)} \le \sigma \le \pi (\alpha_{(1)} + \alpha_{(2)}),\cr
\sigma + \pi \alpha_{(1)} & - \pi (\alpha_{(1)} + \alpha_{(2)}) \le \sigma
\le - \pi \alpha_{(1)},} \right.\cr
\sigma_{(3)} &= - \sigma \qquad\qquad  - \pi(\alpha_{(1)} +
\alpha_{(2)}) \le \sigma
\le \pi (\alpha_{(1)} + \alpha_{(2)}).
}}
In general,
when we write a function of $\sigma$ with a subscript ${(r)}$, it
is to be understood that the function has support only
for $\sigma$ within the range which coincides with string $r$.
For example,
$p_{(r)}(\sigma)$ will denote $p_{(r)}(\sigma) =
\Theta_{(r)}(\sigma)  p_{(r)}(\sigma_{(r)})$, where
\eqn\aaa{
\Theta_{(1)} = \theta(\pi \alpha_{(1)} - |\sigma|), \qquad \Theta_{(2)}
= \theta(
|\sigma| - \pi \alpha_{(1)}), \qquad \Theta_{(3)} = 1.
}

\newsec{Supergravity Vertex}

In this section we consider the cubic coupling of just the
supergravity modes.
This is interesting for its own sake, but more importantly it serves
to clarify the procedure with a minimum of technical complication.
The generators of the pp-wave superalgebra at $\kappa=0$ are
\eqn\sgalg{\eqalign{
H &= {1 \over \alpha} \left( p^2 + {1 \over 4} (\mu \alpha)^2 x^2\right)
-   \mu \theta \Pi \lambda,\cr
P^I &= p^I,\cr
J^{+I} &= {1 \over 2} \alpha x^I,\cr
Q_a^+ &=  \sqrt{2} \lambda_a,\cr
\overline{Q}{}_a^+ &= {\alpha \over \sqrt{2}}  \theta_a,\cr
Q_{\dot{a}}^- &= {2 \over \alpha} \left[ p^I (\gamma_I \lambda)_{\dot{a}}
- i {\alpha \over 2} \mu x^I (\gamma_I \Pi \lambda)_{\dot{a}}\right],\cr
\overline{Q}{}^-_{\dot{a}} &=  p^I (\gamma_I \theta)_{\dot{a}}
+ i {\alpha \over 2} \mu x^I (\gamma_I \Pi \theta)_{\dot{a}}.
}}
Here $p$, $x$, $\lambda$ and $\theta$
are really the zero-modes $p_0$, $x_0$, $\lambda_0$ and $\vartheta_0$
from section 2, but we will omit the subscript throughout this section.

Since $H_3$, $Q^-_3$ and $\overline{Q}{}^-_3$ commute with the
kinematical generators $P^I$, $J^{+I}$, $Q^+$ and $\overline{Q}{}^+$, we start
by constructing a state $|V\rangle$ which satisfies
\eqn\breqone{
\qquad\qquad~~~~
(\hat{p}_{(1)} + \hat{p}_{(2)} + \hat{p}_{(3)})
|V\rangle = 0,
}
\eqn\breqtwo{
(\alpha_{(1)}
\hat{x}_{(1)} + \alpha_{(2)} \hat{x}_{(2)} + \alpha_{(3)} \hat{x}_{(3)})
|V\rangle = 0,
}
\eqn\freqone{
\qquad\qquad~~~~
(\hat{\lambda}_{(1)} + \hat{\lambda}_{(2)} + \hat{\lambda}_{(3)}) |V\rangle
= 0,
}
\eqn\freqtwo{
~~
(\alpha_{(1)}
\hat{\theta}_{(1)} +
\alpha_{(2)} \hat{\theta}_{(2)} +
\alpha_{(3)} \hat{\theta}_{(3)}) |V\rangle
= 0.
}
The vertex $|V\rangle$ enforces conservation of $P^I$, $J^{+I}$, $Q^+$
and $\overline{Q}{}^+$ and forms the basic building block for constructing
the interaction terms in the dynamical generators, which will take
the form
\eqn\aaa{
|H \rangle = \hat{h} |V\rangle,
\qquad
|Q^{- \dot{a}}\rangle = \hat{q}^{- \dot{a}} |V\rangle, \qquad
|\overline{Q}{}^{- \dot{a}}\rangle = \hat{\overline{q}}
{}^{- \dot{a}} |V\rangle
}
for some operators $\hat{h}$, $\hat{q}^{- \dot{a}}$
and $\hat{\overline{q}}{}^{- \dot{a}}$.
Of course these operators should (anti)-commute with the operators
appearing in \breqone-\freqtwo\ so as not to ruin the conservation
laws.
We will then determine $\hat{h}$, $\hat{q}^{- \dot{a}}$ and
$\hat{\overline{q}}{}^{- \dot{a}}$
by requiring the relations
\eqn\reqrel{
\{ Q^{- \dot{a}}, \overline{Q}{}^{-\dot{b}} \} = 2 \delta^{\dot{a} \dot{b}}
H + {\cal{O}}(\kappa^0), \qquad
\{ Q^{- \dot{a}}, Q^{-\dot{b}} \} = 0, \qquad
\{ \overline{Q}{}^{- \dot{a}}, \overline{Q}{}^{-\dot{b}} \} = 0
}
to hold at ${\cal{O}}(\kappa)$.

\subsec{Bosonic Modes}

In this subsection we construct the operator $E_a$ such that
$E_a |0\rangle$ is annihilated by $\sum \hat{p}_{(r)}$ and
$\sum \alpha_{(r)} \hat{x}_{(r)}$.
To construct such a state we use \vertex.  The momentum eigenstate
$|p\rangle$ may be expressed as $\psi(p)|0\rangle$, where
\eqn\aaa{
\psi(p) \sim
\exp \left[{{
- \ha a^\dagger a^\dagger + {2 \over \sqrt{|\alpha| \mu}}
p a^\dagger - {1 \over |\alpha| \mu} p^2}} \right].
}
The symbol $\sim$ denotes that we are ignoring overall factors.
The bosonic part of the integral \vertex\ is then
\eqn\aaa{
E_a^0 \equiv
\int dp_{(1)} dp_{(2)} dp_{(3)} \psi(p_{(1)}) \psi(p_{(2)}) \psi(p_{(3)})
\delta(p_{(1)} + p_{(2)} + p_{(3)}).
}
This is a simple Gaussian integral, and yields the result
\eqn\aaa{
E_a^0 \sim
\exp \left[  \ha \sum_{r,s = 1}^3 a_{(r)}^\dagger M^{rs}
a_{(s)}^\dagger
\right],
}
where
\eqn\aaa{
M^{rs} = \left(\matrix{ 
\beta + 1 &-\sqrt{-\beta(1+\beta)} &-\sqrt{-\beta} \cr
-\sqrt{-\beta(1+\beta)} &-\beta &-\sqrt{1+\beta} \cr
-\sqrt{-\beta} &-\sqrt{1+\beta} & 0
}
\right)
}
and $\beta =  \alpha_{(1)}/\alpha_{(3)}$.
It is trivial to check directly
that the state $E_a^0 |0\rangle$ satisfies the
desired identities \breqone, \breqtwo.
Additionally, we have the useful fact that
\eqn\ronv{
(\hat{x}_{(r)}
- \hat{x}_{(s)})E_a^0 |0\rangle= 0, \qquad \forall r,s.
}

Now let us ask the question:  is $E_a^0|0\rangle$ the
unique state which satisfies
\breqone,\breqtwo?
Indeed, no:
it is easy to see that any state of the form
$f(\IP) E_a^0 |0\rangle$, where $f$ is an arbitrary function of
\eqn\pdef{
\IP = \alpha_{(1)} \hat{p}_{(2)} - \alpha_{(2)} \hat{p}_{(1)},
}
will satisfy the same identities by virtue of the fact that
$\IP$ commutes with $\sum \hat{p}_{(r)}$ and $\sum \alpha_{(r)}
\hat{x}_{(r)}$.  Note that we could have defined $\IP$ with operators
from any two of the three strings (instead
of strings 1 and 2 as we have chosen), but it is easy to see from
\breqone\ and $\sum \alpha_{(r)} = 0$ that they are all 
equivalent when acting on $E_a^0 |0\rangle$.

\subsec{Fermionic Modes}

Conservation of $Q^+_a$ and $\overline{Q}{}^+_a$ implies that the
state $|V\rangle$ must include an operator $E_b^0$ which satisfies
the identities
$\sum \hat{\lambda}_{(r)} E_b^0|0\rangle = 0$ and
$\sum \alpha_{(r)} \hat{\theta}_{(r)} E_b^0|0\rangle = 0$.
These requirements almost uniquely determine
\eqn\ebdef{
E_b^0 \sim {1 \over 8!}\epsilon_{a_1 \cdots a_8}
\hat{\lambda}^{a_1} \cdots \hat{\lambda}^{a_8}|0\rangle, \qquad
{\rm where}~\hat{\lambda} =
\hat{\lambda}_{(1)}+\hat{\lambda}_{(2)}+\hat{\lambda}_{(3)}.
}
This choice also satisfies the useful relation
\eqn\thetaonv{
(\hat{\theta}_{(r)} - \hat{\theta}_{(s)}) E_b^0|0\rangle = 0, \qquad
\forall r,s.
}
The meaning of `almost uniquely' is that we
could include an arbitrary function $f(\Lambda)$ of
\eqn\aaa{
\Lambda = \alpha_{(1)} \hat{\lambda}_{(2)}
- \alpha_{(2)} \hat{\lambda}_{(1)}.
}
This does not spoil
\freqone, \freqtwo\ since $\Lambda$
anticommutes with $\sum \hat{\lambda}_{(r)}$ and
$\sum \alpha_{(r)} \hat{\theta}_{(r)}$. Again the choice
of which two strings to single out is irrelevant.

\subsec{Prefactor}

We have determined that the kinematical constraints
\breqone-\freqtwo\ are satisfied by the state
\eqn\aaa{
|V\rangle \equiv  E_a^0 E_b^0 |0\rangle,
}
and more generally by any state of the form $f(\IP, \Lambda)|V\rangle$
for any function $f$.
Motivated by the expectation (explained in detail
in section 6) that the prefactor will take essentially
the same form as in flat space, we make the ansatz
\eqn\ansatz{\eqalign{
|H \rangle &= \IP^I \IP^J
v_{IJ}(\Lambda) |V \rangle,\cr
|Q^-_{\dot{a}} \rangle &= \IP^I s_{I \dot{a}}(\Lambda) |V \rangle,
\cr
|\overline{Q}{}^-_{\dot{a}} \rangle &= \IP^I t_{I \dot{a}}(\Lambda) |V\rangle,
}}
where $v$, $s$ and $t$ are functions to be determined.
Because of the terms proportional to $\mu x$ in the supercharges
\sgalg,
one might be tempted to include terms with $\mu \IR^I$, where
\eqn\aaa{
\IR \equiv {1 \over \alpha_{(3)}}( x_{(1)}-x_{(2)}),
}
but $\IR|V\rangle = 0$ by virtue of \ronv.
The ${\cal{O}}(\kappa)$ terms in the relations \reqrel\ require
\eqn\reqalg{\eqalign{
\sum_{r=1}^3 Q^{- \dot{a}}_{(r)}
| \overline{Q}{}^{- \dot{b}}\rangle 
+ \sum_{r=1}^3 \overline{Q}{}^{- \dot{b}}_{(r)}
| Q^{- \dot{a}}\rangle 
&= 2 \delta^{\dot{a} \dot{b}} |H\rangle,\cr
\sum_{r=1}^3 Q_{(r)}^{- \dot{a}} | Q^{-\dot{b}}\rangle
+ (\dot{a}\leftrightarrow \dot{b}) &= 0,\cr
\sum_{r=1}^3 \overline{Q}{}_{(r)}^{- \dot{a}} | \overline{Q}{}^{-\dot{b}}\rangle
+ (\dot{a}\leftrightarrow \dot{b}) &= 0.
}}

Let us note the useful relations
\eqn\useful{\eqalign{
\sum_{r=1}^3 [ Q^{- \dot{a}}_{(r)}, \IP^I ]
&=   \mu (\gamma^I \Pi \Lambda)^{\dot{\alpha}},\cr
\sum_{r=1}^3 [ \overline{Q}{}^{- \dot{a}}_{(r)}, \IP^I ] &= \ha \alpha \mu (
\gamma^I \Pi \Theta)^{\dot{a}},\cr
\sum_{r=1}^3 \{ Q^{- \dot{a}}_{(r)}, \Lambda^b \} &=0,\cr
\sum_{r=1}^3 \{ \overline{Q}{}^{- \dot{a}}_{(r)}, \Lambda^b \} &=
 \IP^I (\gamma_I)^{\dot{a} b} - {i \over 2} \alpha \mu \IR^I
(\gamma_I \Pi)^{\dot{a} b},
}}
where $\alpha \equiv \alpha_{(1)} \alpha_{(2)} \alpha_{(3)}$ and
\eqn\aaa{
\Theta = {1 \over \alpha_{(3)}}( \theta_{(1)}
-\theta_{(2)}).
}
Using \breqone-\freqtwo\ and $\sum \alpha_r = 0$, it is easy to see that
we also have
\eqn\vacact{
\sum_{r=1}^3 \overline{Q}{}^{- \dot{a}}_{(r)} |V \rangle = 0,\qquad
\sum_{r=1}^3 Q^{- \dot{a}}_{(r)} |V \rangle = -{2 \over \alpha}
\IP^I (\gamma_I \Lambda)^{\dot{a}} |V\rangle.
}
Note that $\Theta|V\rangle = 0$ as a consequence
of \thetaonv.
Using these relations, we can substitute the ansatz \ansatz\ into
\reqalg\ to obtain equations on $v$, $s$ and $t$.

Three of the equations are identical to the analogous equations
in flat space:
\eqn\eqone{\eqalign{
2 \delta_{\dot{a} \dot{b}} v_{IJ} &= 
\left[\ha (\gamma_I)_{\dot{a} b} {\p s_{J \dot{b}} \over \p \Lambda^b}
+ {1 \over \alpha} t_{I \dot{a}} (\gamma_J \Lambda)_{\dot{b}}
\right]
+ (I \leftrightarrow J),\cr
0 &=\left[
(\gamma_I \Lambda)_{\dot{a}} s_{J \dot{b}}
+ (\dot{a} \leftrightarrow \dot{b})\right] + (I \leftrightarrow
J),\cr
0 &=\left[
(\gamma_I)_{\dot{a} a} {\p t_{J \dot{b}}
\over \p \Lambda^a} + (\dot{a} \leftrightarrow \dot{b})
\right]+ (I \leftrightarrow
J).
}}
These equations arise as coefficients of $\IP^I \IP^J$
terms in \reqalg\ (hence the symmetrization on the $I,J$ indices).
But there are additional equations which come from the terms
in \useful\ proportional to $\mu$.  These terms do not multiply
$\IP^I \IP^J$ so they must vanish separately:
\eqn\eqtwo{
\eqalign{
0 &= (\gamma^I \Pi \Lambda)_{\dot{a}} t_{I \dot{b}} + {1 \over 2}
\alpha (\gamma^I \Pi)_{\dot{b} b} {\p s_{I \dot{a}} \over \p \Lambda_b},\cr
0 &= (\gamma^I \Pi \Lambda)_{\dot{a}} s_{I \dot{b}} + (\dot{a}
\leftrightarrow \dot{b}),\cr
0 &= (\gamma^I \Pi)_{\dot{a} a} {\p t_{I \dot{b}}
\over \p \Lambda^b} + (\dot{a} \leftrightarrow \dot{b}).
}}
The solution to the flat-space equations \eqone\ is \GSB\ 
\eqn\vst{\eqalign{
v_{IJ} &= \delta^{IJ} + {1 \over 6 \alpha^2}
\gamma^{IK}_{ab} \gamma^{JK}_{cd}
\Lambda^a \Lambda^b \Lambda^c \Lambda^d + {16 \over 8! \alpha^4}
\delta^{IJ} \epsilon^{abcdefgh}
\Lambda^a \Lambda^b \Lambda^c \Lambda^d\Lambda^e \Lambda^f \Lambda^g
\Lambda^h,\cr
s_{I\dot{a}} &= 2 \gamma^I_{\dot{a} a} \Lambda^a - {8 \over 6!
\alpha^2} \gamma^{IJ}_{ab} \gamma^J_{c \dot{a}} \epsilon^{abcdefgh}
\Lambda^d\Lambda^e \Lambda^f \Lambda^g \Lambda^h,\cr
t_{I \dot{a}} &= -{2 \over 3 \alpha}
\gamma^{IJ}_{ab} \gamma^J_{c \dot{a}} \Lambda^a \Lambda^b \Lambda^c
- {16 \over 7! \alpha^3} \gamma^I_{a \dot{a}} \epsilon^{abcdefgh}
\Lambda^b \Lambda^c \Lambda^d\Lambda^e \Lambda^f \Lambda^g \Lambda^h.
}}
It is straightforward, though tedious, to verify that the functions
satisfy the additional constraints \eqtwo.

This might seem like a small miracle.
How can the flat space expressions \vst, which are
determined uniquely by super-Poincar\'e invariance, handle the
additional burden of satisfying the extra
constraints \eqtwo?  It turns out that \vst\ would satisfy \eqtwo\ for
any matrix $\Pi$ which is symmetric and traceless.
Together with $\Pi^2 = 1$, which is required to satisfy
$\{ Q^-, \overline{Q}{}^- \} = 2 H$ at the level of free fields, we see that this
miracle works {\it only} for the matrix $\Pi = {\rm diag}(1_4,
-1_4)$ (up to a change of basis),
which is fortunately the $\Pi$ that we happen
to be interested in.
Once again, we see how the pp-wave superalgebra miraculously
makes our calculation almost as easy as in flat space.

In flat space supergravity, one is
free to drop all but the quartic term in $v$, the quintic term
in $s$, and the cubic term in $t$ for on-shell three-particle
amplitudes (see \GreenTK).
To see why this is so for $v$, note that the first and last terms
are proportional to $\delta^{IJ}$ and hence give
a contribution like
\eqn\vanishes{
|H \rangle \sim \IP^2 |V\rangle.
}
It is easy to check that
\eqn\aaa{
\IP^2|V\rangle = - \alpha \sum_{r=1}^3 H_{(r)}|V\rangle,
}
so in fact the matrix element of \vanishes\ with any three on-shell
states satisfying energy conservation will vanish.
In fact the same is true for the states we have constructed
in the pp-wave background, since here
\eqn\aaa{
\IP^2|V\rangle = - \alpha \sum_{r=1}^3\left[
H_{(r)} - {1 \over 4} \mu^2 \alpha_{(r)}
\hat{x}_{(r)}^2 -  \mu \hat{\lambda}_{(r)}
\Pi \hat{\theta}_{(r)}
\right]|V\rangle.
}
The second term vanishes when acting on $|V\rangle$ as a consequence of
\ronv\ and \breqtwo, and the third term vanishes when acting on
$|V\rangle$ as a consequence of \freqone\ and \thetaonv.
We conclude that $\IP^2 |V \rangle$ 
vanishes in on-shell matrix elements.

Finally, let us rewrite $|H\rangle$ in a form which will facilitate
the inclusion of string modes. 
Let us define $\IK$ to include only the raising operators in
$\IP$,
\eqn\ikdef{
\IK \equiv 2 \left. \IP \right|_{\rm raising}
= - {\sqrt{\mu}}
( \sqrt{\alpha_{(1)}} \alpha_{(2)} a_{(1)}^\dagger
- \alpha_{(1)} \sqrt{\alpha_{(2)}} a_{(2)}^\dagger)
=   \IP - {i \over 2} \mu \alpha \IR.
}
Since $\IR$ annihilates $|V\rangle$, can express
the result
\ansatz\ in the equivalent form
\eqn\anssugra{
\eqalign{
|H\rangle &=  \left(
\IK^I \IK^J - \ha \mu \alpha \delta^{IJ}\right) v_{IJ}(\Lambda)
|V\rangle,\cr
|Q_{\dot{a}}^-\rangle &=  \IK^I s_{I \dot{a}}(\Lambda)|V\rangle,
\cr
|\overline{Q}{}_{\dot{a}}^-\rangle &=  \IK^I t_{I \dot{a}}(\Lambda)|V\rangle,
}
}
where the trace term in $|H\rangle$ came from
\eqn\aaa{
[ \IR^I, \IP^J] = i \delta^{IJ}.
}
In the representation \anssugra\ it is not manifest
(though of course still true) that the
terms in $v^{IJ}$ proportional to $\delta^{IJ}$ give zero contribution
when the three external states are on-shell.

\newsec{String Theory Vertex}

In this section we determine
the operators $E_a$ and $E_b$, constructed out of all the
string modes,
such that the vertex $|V \rangle = E_a E_b|0\rangle$
satisfies the kinematic constraints
\eqn\aaa{
\sum_{r=1}^3 P_{(r)}(\sigma) |V\rangle = \sum_{r=1}^3
e(\alpha_{(r)}) X_{(r)}(\sigma) |V\rangle =
\sum_{r=1}^3 \Lambda_{(r)}(\sigma) |V\rangle =
\sum_{r=1}^3 e(\alpha_{(r)}) \Theta_{(r)}(\sigma)|V\rangle = 0.
}

\subsec{Bosonic Modes}

The three-string vertex contains a $\Delta$-functional of
momentum conservation on the worldsheet:
\eqn\Deltapone{
\Delta \left[ \sum_{r=1}^3 p_{(r)}(\sigma) \right].
}
Delta-functionals of this form can be defined as an infinite
product of delta functions for the individual Fourier modes of the argument:
\eqn\aaa{
\Delta[ f(\sigma)] = \prod_{m=-\infty}^\infty \delta
\left( \int_{0}^{2 \pi |\alpha_{(3)}|} e^{i
m \sigma/|\alpha_{(3)}|} f(\sigma) \ d\sigma \right)
}
The Fourier modes of \Deltapone\ were worked out in
\GSB, and it is
an easy matter to use \transcribe\ to transcribe their result into
the basis \xandp\ for $p(\sigma)$ used in this paper.
The result is
\eqn\Deltap{
\Delta \left[ \sum_{r=1}^3 p_{(r)}(\sigma) \right]
= \prod_{m=-\infty}^\infty \delta \left(
p_{m(3)} + \sum_{n=-\infty}^\infty(X^{(1)}_{mn} p_{n(1)} + X^{(2)}_{mn}
p_{n(2)})\right).
}
The matrices $X^{(1)}$ and $X^{(2)}$ are written in appendix A (we
define $X^{(3)} = {\bf 1}$).
The $\Delta$-functional \Deltap\  determines
the Fourier modes of the third string  in terms of the modes of the 
other two strings.

The properly normalized wavefunctions for momentum eigenstates
in terms of the raising operators are
\eqn\wavfun{
\psi(p_{n(r)}) = \left( {2 \over \pi \omega_{n(r)}}\right)^{1/4} \exp
\left[
- {1 \over \omega_{n(r)}} p_{n(r)}^2
+ {2 \over \sqrt{\omega_{n(r)}}} p_{n(r)} a_{n(r)}^\dagger
- {1 \over 2} a_{n(r)}^\dagger a_{n(r)}^\dagger
\right]
}
Therefore we have to do
the integral
\eqn\eastring{
E_a=
\int\left(\prod_{r=1}^3 \prod_{n=-\infty}^\infty
d p_{n(r)} \psi(p_{n(r)})
\right) \Delta \left[ \sum_{r=1}^3 p_{(r)}(\sigma)\right].
}
This is a fairly straightforward Gaussian integral, and in appendix
B we show that the result is
(up to an overall factor which is computed in the appendix)
\eqn\bvertex{
E_a \sim
\exp \left[  {1 \over 2} \sum_{r,s=1}^3 \sum_{m,n=-\infty}^\infty
a_{m (r)}^\dagger \overline{N}^{(rs)}_{mn} a_{n(s)}^{\dagger}
\right] |0\rangle,
}
where the Neumann matrices $\overline{N}^{(rs)}_{mn}$ are given by
\eqn\neuman{
\overline{N}_{mn}^{(rs)} = \delta^{rs} \delta_{mn} - 2 \sqrt{
\omega_{m(r)} \omega_{n(s)}
}
(X^{(r) {\rm T}} \Gamma_a^{-1}
X^{(s)})_{mn},
}
in terms of the matrix
\eqn\gammadef{
(\Gamma_a)_{mn} = \sum_{r=1}^3 \sum_{p=-\infty}^\infty \omega_{p(r)}
X^{(r)}_{mp} X^{(r)}_{np}.
}

It is straightforward to check that the result
\neuman\ reduces to the flat space Neumann matrices as $\mu \to 0$,
although the zero modes must be separated and treated more carefully
to find their contribution.  In flat space, the worldsheet theory
is conformal in light-cone gauge, so that it is natural
to use a basis of left- and right-moving oscillators,
${\alpha}_n = \sqrt{n/2}(a_n -i a_{-n})$
and $\tilde{\alpha}_n = \sqrt{n/2}(a_n + i a_{-n})$ (for $n>0$).
The fact that these decouple in a conformal theory implies
an identity which in our notation reads
\eqn\decoupl{
\overline{N}^{(rs)}_{mn} =-\overline{N}^{(rs)}_{-m,-n}.
}
In the notation of \GSB\ the corresponding identity looks less trivial,
but it is not hard to show that it is true for $\mu=0$.
However, it is easy to check that the relation \decoupl\ no longer holds when
$\mu \ne 0$, as expected for a worldsheet theory that is
not conformal.

\subsec{Fermionic Modes}

The three-string vertex contains a $\Delta$-functional of
fermionic momentum conservation on the worldsheet:
\eqn\Deltalambda{
\Delta \left[ \sum_{r=1}^3 \lambda_{(r)}(\sigma) \right] =
\prod_{m=-\infty}^\infty \delta
\left(\sum_{r=1}^3 \sum_{n=-\infty}^\infty
X^{(r)}_{mn} \lambda_{n(r)}
\right).
}
It is natural to separate the zero mode, since the Grassmann
wavefunction will couple the $n>0$ and $n<0$ modes.
The zero modes give a contribution of $E_b^0$ given by
\ebdef.

For the rest of this section, we always take $n>0$ and write $-n$
to be explicit when we mean a negative component.
In order to write \Deltalambda\ in an oscillator basis, we need to write
the fermionic analogue of the wavefunction \wavfun.
Consider a state of the form
\eqn\fwavfun{
|\lambda_n \lambda_{-n} \rangle 
\sim
\exp\left[
{2 
\over \alpha}\left(
 \lambda_{-n} \tb_n^\dagger
-  \lambda_n \tb_{-n}^\dagger
+{n \over 2\omega_n}
\tb_n^\dagger
\tb_{-n}^\dagger
\right)
\right]
|0\rangle,
}
where we define
\eqn\aaa{
\tb^\dagger_{\pm n} = 
\sqrt{|\alpha|} {(1 + \rho_n \Pi) \over
c_n(1 - \rho_n^2)} b_{\pm n}^\dagger.
}
It is easy to check that
this state satisfies the eigenvalue equations
\eqn\aaa{\eqalign{
{\sqrt{|\alpha|} \over 2} c_n \left[ e(\alpha) (1 - \rho_n \Pi) b_{-n}
+  (1 + \rho_n \Pi) b_n^\dagger \right] |\lambda_n \lambda_{-n}\rangle
&= \lambda_n |\lambda_n \lambda_{-n}\rangle,\cr
 {\sqrt{|\alpha|} \over 2} c_n \left[
- e(\alpha)
(1 - \rho_n \Pi) b_n + (1 + \rho_n \Pi) b_{-n}^\dagger \right]
|\lambda_n \lambda_{-n}\rangle
&= \lambda_{-n} |\lambda_n \lambda_{-n}\rangle.
}}
Therefore, 
\fwavfun\ is the desired expression, up to a term without
$b^\dagger$ operators.
That term is determined by checking
that $\langle 0|\lambda_n \lambda_{-n}\rangle$ is the ground
state wavefunction for the $b_n$ and $b_{-n}$ harmonic oscillators.
To determine this wavefunction we invert the transformation
\iofshe\ to find
\eqn\aaa{
b_n =  {\sqrt{|\alpha|} \over 2}
c_n \left[ (1  + \rho_n \Pi) {\p \over \p \lambda_n}
-  {2 \over \alpha} (1 - \rho_n \Pi) \lambda_{-n} \right],
}
where we have represented $\hat{\vartheta}_n = {\p \over \p \lambda_n}$.
The ground state wavefunction should be annihilated by $b_n$, which determines
\eqn\aaa{
\langle 0 | \lambda_n \lambda_{-n} \rangle = 
 \exp \left[
 {2 \over \alpha} \lambda_n P^2_n \lambda_{-n}\right], \qquad
P_n = 
{(1 - \rho_n \Pi) \over \sqrt{1 - \rho_n^2}}.
}
We have omitted the overall normalization.
Putting everything together gives the desired expression
$|\lambda_n \lambda_{-n}\rangle = \chi(\lambda_n, \lambda_{-n})|0\rangle$,
where
\eqn\aaa{
\chi(\lambda_n, \lambda_{-n}) =  \exp \left[
{2 \over \alpha} \left(
 \lambda_n P^2_n \lambda_{-n} + \lambda_{-n}
\tb_n^\dagger - \lambda_n \tb_{-n}^\dagger + {n \over 2 \omega_n}
\tb_n^\dagger
 \tb_{-n}^\dagger\right)\right].
}

The integral we are interested in,
\eqn\aaa{
E_b = \int \left( \prod_{r=1}^3 \prod_{n=1}^\infty  d\lambda_{n(r)}
d\lambda_{-n(r)} \ \chi(\lambda_{n(r)}, \lambda_{-n(r)})\right)
\Delta \left[ \sum_{r=1}^3 \lambda_{(r)}(\sigma)\right],
}
can be performed in much the same way as the bosonic
integral in appendix A.
To express the result we introduce the matrices
\eqn\aaa{
Y^{(r)}_{mn} = P_{|m|}^{(3)} X^{(r)}_{mn} P_{|n|}^{(r)-1},
\qquad
\overline{Y}^{(r)}_{mn} = \alpha_{(r)}
P_{|m|}^{(3)} X^{(r)}_{-m, -n} P_{|n|}^{(r)-1}
}
with the understanding that $P_0 = 1$,
and
\eqn\aaa{
\Gamma_b = \sum_{r=1}^3 {Y}^{(r)} \overline{Y}^{(r) {\rm T}}.
}
Then the vertex takes the form\foot{There is, in addition,
a coupling of the zero mode $\Lambda$ to the non-zero modes
$b_{m}^\dagger$ which we have not written.  It can be obtained
by careful consideration of the $m=0$ limit of $Q_{mn}$.}
\eqn\ebstring{
E_b=
\exp \left[ \sum_{r,s=1}^3 \sum_{m,n=1}^\infty
 b^\dagger_{-m (r)}
Q^{(rs)}_{mn}
b^\dagger_{n(s)}
\right] E_b^0 |0\rangle,
}
where we have restored the zero mode contribution from \ebdef, and
where
\eqn\aaa{
Q^{(rs)}_{mn} =
{1 \over \alpha_{(r)}}
 V^{(r)}_m \left[ \delta^{rs} \delta_{mn}
\left(1 + {\mu \alpha_{(r)} \over \omega_{n(r)}} \Pi
\right)
- 2(\overline{Y}^{(r) {\rm T}} \Gamma_b^{-1} Y^{(s)} )_{mn}
\right] V^{(s)}_n,
}
with
\eqn\aaa{
V^{(r)}_n = \sqrt{|\alpha_{(r)}|}
{ (1 + \rho_{n(r)} \Pi)^2 \over c_{n(r)} (1 - \rho_{n(r)}^2)^{3/2}}.
}

\newsec{Interaction Point Operator}

In this section we derive the operators $\hat{h}_3$,
$\hat{q}_3$ and $\hat{\overline{q}}_3$ which must be inserted in the
three-string vertex in order to preserve supersymmetry.
The necessity of introducing these operators arises
from a short-distance effect on the worldsheet.  Since this
short-distance behaviour is unaffected by the addition
of a mass term $\mu$ to the worldsheet action,
the interaction point operators should be the same for the pp-wave
as in flat space.  This expectation has also been noted in
\BerkovitsZV.

\subsec{The Necessity of a Prefactor}

Suppose\foot{This argument
appears in \GSW.}
we tried to take $\hat{h}_3 = 1$ so that $|H_3\rangle = |V\rangle$.
Then the ${\cal{O}}(\kappa)$ terms in the relation
$[ Q^-, H] = 0$
tell us that
\eqn\aaa{
0 =
\sum_{r=1}^3 H_{(r)} |Q^-\rangle + \sum_{r=1}^3 Q^{-}_{(r)} |V \rangle.
}
Consider the matrix element of this equation between on-shell
external states, so that $\sum_r H_r = 0$.
We need then $\sum Q^-_{(r)} |V \rangle = 0$.  From
\qminus, this means
we would need
\eqn\sing{
\sum_{r=1}^3  \int d\sigma_{(r)}
\left[ 4 \pi e(\alpha_{(r)}) p_{(r)}^I
\gamma_I \lambda_{(r)} - {i \over 4 \pi} \p_\sigma
x_{(r)}^I\gamma_I
\theta_{(r)} - i \mu x_{(r)}^I \gamma_I \Pi \lambda_{(r)}\right]
|V\rangle = 0.
}
Naively, the kinematical constraints \breqone-\freqtwo\ indeed guarantee
that this is zero.
The problem is that the operators in \sing\ applied
to $|V\rangle$ are singular near the
interaction point $\sigma = \pi \alpha_{(1)}$.  For example,
we have $p(\sigma) \lambda(\sigma) |V \rangle
\sim \p_\sigma x(\sigma)\theta(\sigma) |V \rangle
\sim \epsilon^{-1} |V\rangle$ for $\sigma
= \pi \alpha_{(1)}-\epsilon$.  Therefore, although \sing\ vanishes
pointwise
for all $\sigma \ne \pi \alpha_{(1)}$, the singular operators
nevertheless give
a finite contribution when integrated over $\sigma$.  This
contribution can be calculated by deforming the contour in an appropriate
way and reading off the residues of the poles \GSW.

However, the last term in \sing\ is not singular:
$x(\sigma) \lambda(\sigma)|V\rangle \sim \epsilon^0 |V\rangle$.
Therefore \sing\ gives precisely the same constraint as in
flat space, which suggests
that the form of the prefactor is the same
as it is in flat space.
This is a manifestation of the aforementioned fact that
$\mu$ should not affect the short-distance behaviour of the
worldsheet theory.
Armed with this knowledge,
all that remains is to find stringy
generalizations of the operators $\IP$ and $\Lambda$ which we
used in section 4 to write the prefactor in supergravity.
These should be chosen so that when we
make an ansatz analogous to \ansatz, the form of the
equations \eqone\ and \eqtwo\ on the functions
$s,t$ and $v$ remains unchanged.
Then the solutions $s, t$ and $v$ will also be
unchanged.\foot{They will change slightly when we go to
string theory because we will need terms which are not
symmetric in $I,J$.
By ``unchanged,'' we mean
unchanged relative to flat space string theory, where
such terms already appear.}

Now the derivation of the equations \eqone\ and \eqtwo\ follows
quite trivially from the identities \useful\ and \vacact.
Therefore, if we want to preserve these equations, we should
look for stringy generalizations of $\IP$ and $\Lambda$
which preserve the relations \useful\ and \vacact\ as closely
as possible.
For example, from the second equation in \vacact, we see that
we can read off $\IP$ and $\Lambda$ by considering the action
of $\sum Q^-_{(r)}$ on $|V\rangle$.
But we have already considered this in \sing, and concluded that
the result is the same as in flat space!
Therefore, we conclude that the stringy generalizations of
$\IP$ and $\Lambda$ are {\it identical} to the corresponding
flat space expressions in the continuum basis.
These factors are written in the next subsection.

\subsec{Full Vertex with Prefactor}

A careful analysis of the singular terms in \sing, which are
the same as in flat space, necessitate consideration
of the operators defined by \GSB
\eqn\limits{\eqalign{
P|V\rangle &= \lim_{\sigma \to \pi \alpha_{(1)}}
-2 \pi \sqrt{-\alpha} (\pi \alpha_{(1)} - \sigma)^{1/2}
\left( P_{(1)}(\sigma) + P_{(1)}(-\sigma)\right)|V\rangle,\cr
\p X|V\rangle &= \lim_{\sigma \to \pi \alpha_{(1)}}
-2 \pi \sqrt{-\alpha} (\pi \alpha_{(1)} - \sigma)^{1/2}
\left(
\p_\sigma X_{(1)}(\sigma) + \p_\sigma X_{(1)}(-\sigma) 
\right)|V\rangle,\cr
\Lambda|V\rangle &= \lim_{\sigma \to \pi \alpha_{(1)}}
-2 \pi \sqrt{-\alpha} (\pi \alpha_{(1)} - \sigma)^{1/2}
\left( \Lambda_{(1)}(\sigma) + \Lambda_{(1)}(-\sigma)\right)|V\rangle.
}}
Although we have singled out string 1, the final result turns out
to be independent of this choice.

In terms of these, the operators which appear
in the cubic string vertex are\foot{Note that
$P$ and $\p X$ only involve creation operators,
so these are stringy generalizations of $\IK$, rather
than of $\IP$, from  section 4.  Therefore we take
the prefactor as written in the form \anssugra\ rather than \ansatz.}
\eqn\svtstring{\eqalign{
\hat{h}_3 &= \left[\left( P^I + {\textstyle{1 \over 4 \pi}} \partial X^I \right)
\left( P^J - {\textstyle{1 \over 4 \pi}} \p X^J \right) v_{IJ}(\Lambda)
- \half \mu \alpha \delta^{IJ}
\right],\cr
\hat{q}_3^{- \dot{a}} &=
P^I s_{I \dot{a}}(\Lambda) + {i \over 4 \pi} \p X^I t_{I \dot{a}}(\Lambda),\cr
\hat{\overline{q}}_3^{-
\dot{a}} &=  P^I t_{I \dot{a}}(\Lambda) - {i \over 4 \pi}
\p X^I s_{I \dot{a}}(\Lambda).
}}
The functions $s$ and $t$ are
as given in \vst, but the function $v$ has some new terms which
are not symmetric in $I,J$ and therefore were ignored when
we were working just with the supergravity modes.
The full expression is
\eqn\aaa{
\eqalign{
v^{IJ}&= \delta^{IJ} - {i \over \alpha} \gamma^{IJ}_{ab}
\Lambda^a \Lambda^b + {1 \over 6 \alpha^2}
\gamma^{IK}_{ab} \gamma^{JK}_{cd} \Lambda^a \Lambda^b \Lambda^c
\Lambda^d\cr
&~- {4 i\over 6! \alpha^3} \gamma^{IJ}_{ab}
\epsilon_{abcdefgh}\Lambda^c\Lambda^d\Lambda^e \Lambda^f
\Lambda^g \Lambda^h
+ {16 \over 8! \alpha^4} \delta^{IJ}
\epsilon_{abcdefgh}
\Lambda^a \Lambda^b \Lambda^c\Lambda^d\Lambda^e \Lambda^f
\Lambda^g \Lambda^h.
}}

Since the result \svtstring\ automatically satisfies
the superalgebra  at $\mu = 0$, we need only check
the ${\cal{O}}(\mu)$ terms in \reqalg.
The generalization of the relations \useful\ to include string modes is
\eqn\usefulstring{
\eqalign{
[ Q^{- \dot{a}}_{(r)}, P^I] &= 
- {1 \over 4 \pi} (\gamma^I \p \Theta)^{\dot{a}}
+ \mu (\gamma^I \Pi \Lambda)^{\dot{a}},\cr
\{ Q^{-\dot{a}}_{(r)}, \Lambda^b\} &= - {i \over 4 \pi}
\p X^I (\gamma_I)^{\dot{a} b},\cr
[ Q^{-\dot{a}}_{(r)}, \p X^I] &= - 4 \pi i (\gamma^I \p
\Lambda)^{\dot{a}},\cr
[ \overline{Q}{}^{-\dot{a}}_{(r)}, P^I ] &= (\gamma^I \p \Lambda)^{\dot{a}}
- {1 \over 4 \pi} \mu  (\gamma^I \Pi \Theta)^{\dot{a}},\cr
\{ \overline{Q}{}^{-\dot{a}}_{(r)}, \Lambda^b\} &= P^I (\gamma_I)^{\dot{a} b}
+ {i \over 4 \pi} \mu X^I (\gamma_I \Pi)^{\dot{a} b},\cr
[ \overline{Q}{}^{- \dot{a}}_{(r)}, \p X^I ] &= -i (\gamma^I \p
\Theta)^{\dot{a}},
}}
while the generalization of \vacact\ is
\eqn\aajd{\eqalign{
Q^{- \dot{a}} |V\rangle &= - {2 \over \alpha} P^I
(\gamma_I
\Lambda)^{\dot{a}} |V\rangle,\cr
\overline{Q}{}^{-\dot{a}}|V\rangle &= 
-{i \over 2 \pi \alpha} \p X^I (\gamma_I \Lambda)^{\dot{a}}
|V\rangle.
}
}

The operators $X$, $\Theta$, $\p \Lambda$ and $\p \Theta$ are
defined via the same limiting process of \limits.
Then $X|V\rangle = \Theta|V\rangle = 0$,
while the resulting operators $\p \Lambda$ are $\p \Theta$ are singular
when acting on $|V\rangle$.
To render all of the relations \usefulstring\ finite
one should always work with the left- or right-moving
linear combinations $Q^- \pm i \overline{Q}{}^-$ and $P \pm {1 \over 4 \pi} \p X$, in
which case all singular terms cancel.
However we do not need to worry about this since we
are only interested in the
${\cal{O}}(\mu)$ terms in \usefulstring---we know from \GSB\ that
the terms independent of $\mu$ work out properly.

Using  \usefulstring\ and \aajd, it is trivial to check that the
${\cal{O}}(\mu)$ constraints
in \reqalg\ reduce to the equations \eqtwo\ which we have
already shown to be satisfied by $s$ and $t$.
This completes the proof that the prefactors 
\svtstring\ are essentially the same as in flat space, when written in a
continuum basis.  It would be straightforward to express
these operators in the number basis by acting with the operators
\limits\ on the bosonic and fermionic Neumann matrices in $|V\rangle$.

\newsec{String Bits and a $1/\mu p^+$ Expansion}

One of the motivations for this work is the hope
to compare string interactions
in the pp-wave background with predictions from the field theory
of \BMN, which is expected to be perturbative in
$1/\mu p^+$ \seven\ (we still
have $\alpha'=2$).  Since we have obtained the Neumann matrices
for arbitrary $\mu p^+$, it is therefore desirable to develop an
expansion in $1/\mu p^+$ for large $\mu$.
In this section we outline how this might be accomplished.
We hope to address this in more detail in
future work.

In the limit of large $\mu p^+$,
all Fourier modes of the string have the same
energy, $\omega = |\alpha| \mu$, so the Fourier basis is no longer
particularly natural.
Instead of having a harmonic oscillator for each Fourier mode of the
string, it is more natural to have a harmonic oscillator for each
point on the worldsheet.  This is because in this limit,
the worldsheet action for the string becomes
ultra-local, so that there is no coupling between $x(\sigma)$ and
$x(\sigma')$ for $\sigma \ne \sigma'$.\foot{We are grateful
to L. Motl for emphasizing this point to us.}
Consideration of the dual field theory also suggests that it is
natural to work in a local basis on the worldsheet, with one
harmonic oscillator for each string `bit' of \BMN.

With this point of view in mind we consider the case when
$\alpha \ne 0$ is an integer, and
we define $J = |\alpha| = 2 |p^+|$.\foot{The precise
relation between $p^+$ in string theory and $J$ in 
the gauge theory is $J = \mu p^+ \alpha' \sqrt{\lambda}$,
where $\lambda$ is the `t Hooft coupling.}
Discretizing the string worldsheet into $J$ bits, we write
the fields $x(\sigma)$ and $p(\sigma)$ (note that we continue to
suppress the transverse index $I=1,\ldots,8$) as
\eqn\coolxandp{
x(\sigma) = \sqrt{J} \sum_{a=0}^{J-1}
{\tt x}_{a} \chi_a(\sigma), \qquad
p(\sigma) = {1 \over 2 \pi \sqrt{J}} \sum_{a=0}^{J-1}
{\tt p}_{a} \chi_a(\sigma),
}
where $\chi_a(\sigma)$ is a step function with support
in the interval $[2 \pi a, 2 \pi (a+1)]$.
The coordinate $\sigma$ is identified modulo $2 \pi J$.
The commutation relations $[X(\sigma), P(\sigma')] = i\delta(\sigma-
\sigma')$ are a consequence of
\eqn\aaa{
[ \hat{\tt x}_a, \hat{\tt p}_b ] = i \delta_{ab}.
}
Substituting \coolxandp\ into
the Hamiltonian \hghd\ and ignoring the $\p_\sigma X$
terms gives
\eqn\coolh{
H = {e(\alpha) \over J} \sum_{a=0}^{J-1}
\left[\hat{\tt p}_a^2 + {1 \over 4} (\mu J)^2 \hat{\tt x}_a^2 \right].
}
In terms of
\eqn\coolladder{
{\tt a}_{a} = {1 \over \sqrt{ \mu J}} \hat{\tt p}_a - {i \over 2}
\sqrt{\mu J} \hat{\tt x}_a,
}
we have
\eqn\aaa{
H = \mu e(\alpha) \sum_{a=0}^{J-1}
{\tt a}_a^\dagger
{\tt a}_a,
}
where we have dropped the normal-ordering constant.

Now consider the interaction of three strings with
$J_3 = J_1 + J_2 \equiv J$.
We define
\eqn\coolxandp{
x_{(r)}(\sigma) = \sqrt{J_r} \sum_{a=0}^{J-1} {\tt x}_{a(r)}
\chi_{a - \ha J_1}(\sigma),\qquad
p_{(r)}(\sigma) = {1 \over 2 \pi \sqrt{J_r}}
\sum_{a=0}^{J-1} {\tt p}_{a(r)}
\chi_{a - \ha J_1}(\sigma).
}
The step functions have been shifted so that $a = 0$ coincides with
$\sigma = - \pi J_1 = - \pi \alpha_{(1)}$.
It is to be understood that
${\tt x}_{a(1)}$ and ${\tt p}_{a(1)}$ are zero for
$a \ge J_1$, while ${\tt x}_{a(2)}$ and ${\tt p}_{a(2)}$
are zero for $a < J_1$.

Now we wish to write down the vertex $|V\rangle$
which expresses the constraint
\eqn\aaa{
\Delta\left[ x_{(1)}(\sigma) + 
x_{(2)}(\sigma)
- x_{(3)}(\sigma)\right].
}
We should do an integral analogous to \eastring, with
wavefunctions $\psi({\tt{p}}_a)$ appropriate for the Hamiltonian
\coolh, but it is trivial to check directly without
having to do the integrals that the solution
is
\eqn\coolv{
|V\rangle = \exp \left[
\sum_{a=0}^{J_1-1} {\tt a}_{a(1)}^\dagger {\tt a}_{a(3)}^\dagger
+ \sum_{a=J_1}^{J-1} {\tt a}_{a(2)}^\dagger {\tt a}_{a(3)}^\dagger
\right] |0\rangle.
}
Writing \coolv\ it in the form
\eqn\aaa{
\exp \left[ {1 \over 2} \sum_{r,s = 1}^3 \sum_{a,b = 0}^{J-1}
{\tt a}_{a(r)}^\dagger \overline{N}_{ab}^{(rs)} {\tt a}_{b(s)}^\dagger \right],
}
shows that the Neumann matrices  $\overline{N}_{ab}^{(rs)}$ are diagonal
in $ab$ and are given by
\eqn\cntwo{
\overline{N}^{(rs)}_{aa} = \left(
\matrix{ 0 & 0 & 1 \cr
0 & 0 & 0\cr
1 & 0 & 0}\right)_{rs}, \qquad
\overline{N}^{(rs)}_{aa} = \left(
\matrix{ 0 & 0 & 0 \cr
0 & 0 & 1\cr
0 & 1 & 0}\right)_{rs},
}
respectively for $a < J_1$ and $a \ge J_1$.

A systematic expansion in $1/\mu p^+$ can be developed by
retaining sub-leading terms in the Hamiltonian 
\coolh.  These would modify the wave functions $\psi({\tt{p}}_a)$
used to calculate the Neumann matrices through the integral \eastring.
The main subtlety in deriving a $1/\mu p^+$ expansion is the fact 
that the limit
$\mu p^+ \to \infty$ does not seem continuous.
Although the modes are degenerate at $\mu p^+ = \infty$, for any
finite value of $\mu p^+$, no matter how large,
there are always string modes with $n \gg
\alpha' \mu p^+$ which have much higher energy than the low-$n$ modes.
Even if one restricts attention to states which are occupied for
small $n$, the calculation of the Neumann matrices requires
the inversion of an infinite matrix $\Gamma$ which is sensitive
to arbitrarily large $n$. 

It is also interesting to consider the change
of basis between
$x_{n(r)}$
and ${\tt x}_{a(r)}$, which takes the form
\eqn\aaa{
x_{n(r)} = \sum_{a=0}^{J-1} T_{na}^{(r)} {\tt x}_{a(r)}
}
for some easily calculated matrices $T_{na}^{(r)}$.
Since an identical formula relates
$p_{n(r)}$ to ${\tt p}_{n(r)}$, we can write
the ladder operators in the number basis as
\eqn\bogo{
a_{n(r)} = \sum_{a=0}^{J-1} T_{na}^{(r)}
\left[ {1 \over \sqrt{\omega_{n(r)}}} {\tt p}_{a(r)}
+ {i \over 2} \sqrt{\omega_{n(r)}} {\tt x}_{a(r)}
\right], \qquad \omega_{n(r)} = \sqrt{n^2 + (\mu J_r)^2}.
}
Comparison with
\coolladder\ shows that
in the strict $\mu J = \infty$ limit, 
the Bogolyubov transformation \bogo\ does not 
mix creation and annihilation operators.
However for finite $\mu$ the Bogolyubov transformation
is nontrivial.
This greatly complicates the calculation of 
factors like
$\exp[(a^\dagger)^2 ] \to \exp[ ({\tt a} + {\tt a}^\dagger)^2]$
which now have exponentials of non-commuting operators.

\newsec{Conclusion}

The bulk of this paper contains detailed calculations of various
pieces of the three-string vertex. We summarize here some of the
more interesting qualitative features of the result.
The normal modes of a string in the pp-wave background
have frequency $\omega_n = \sqrt{n^2 + (\alpha' p^+ \mu)^2}$.
Since even the zero-mode
has nonzero energy, there is no need to separate the modes into
left-movers, right-movers, and the zero modes.  We simply work
with all $n=-\infty,\ldots,\infty$ at once.
The cubic string vertex written in a number oscillator basis
contains a factor
\eqn\answer{
\exp \left[  {1 \over 2} \sum_{r,s=1}^3 \sum_{m,n=-\infty}^\infty
a_{m (r)}^\dagger \overline{N}^{(rs)}_{mn} a_{n(s)}^{\dagger}
\right]|0\rangle,
}
where $r,s$ label the three strings (note that
in our conventions,
$a_m^\dagger$ is a raising operator regardless of the sign of $m$)
and a formula for the Neumann
matrices $\overline{N}^{(rs)}_{mn}$ is presented in \neuman.
The slightly more complicated formula for the fermionic
oscillators is given in \ebstring.
In flat space, the left- and right-movers on the string worldsheet
decouple in the corresponding expression for \answer.  This is
of course no longer true in the present case since the worldsheet
action is not conformal in light-cone gauge.
Note that in this paper we have derived the `un-amputated'
Neumann coefficients, the analogue of $\overline{N}$ in \GSW.
It would be interesting to derive formulas for the `amputated'
Neumann matrices $N$, which might be more useful in studying on-shell
processes.
The vertex also contains a prefactor, given by the operator
\svtstring\ evaluated at the interaction point.
It is easy to check that the form of this operator (written in
the continuum basis as in \limits) is essentially identical to the corresponding
expression in flat space \GSB.

We have not obtained an analytic expression for the matrix
elements $\overline{N}^{(rs)}_{mn}$
for arbitrary $\mu$, because \bvertex\ requires
the inversion of a complicated infinite matrix \gammadef.  In practice, we have
resorted to numerical calculation using a level-truncation scheme.
Of particular interest is the limit $\mu p^+ \to \infty$, since this
is the weak-coupling limit in the dual Yang-Mills theory.
In this limit it is more convenient to use a basis where one
has one harmonic oscillator for each point on the string worldsheet.
The Neumann matrices in this limit are presented in section 7, and we
have discussed how one might develop a $1/\mu p^+$ expansion
in order to compare with transition
amplitudes calculated from the field theory.

A more ambitious project would be to compare one-loop corrections
to the masses of string states in the pp-wave background.
The corresponding field theory calculation has been achieved in
\seven\ for some states.
Reproducing this calculation on the string theory side
would require knowledge of a contact term in the
string field theory Hamiltonian which appears at ${\cal{O}}(\kappa^2)$.
It would also be interesting to study the interactions
of open strings.  D-branes 
in the pp-wave background have recently been considered in
\refs{\BilloFF
\ChuIN
\DabholkarZC
\KumarPS
\BakRQ -
\SkenderisVF}, and open strings have
been constructed in
\refs{\BerensteinZW, \LeeCU}.
Finally, it would be very interesting if the $\mu p^+ = \infty$
string
theory is sufficiently simple to allow an exact solution---meaning
a construction of the interaction Hamiltonian to all orders.

\bigskip
\bigskip
\bigskip

\centerline {\bf Acknowledgements}

\smallskip

We are grateful to M. Headrick, N. Itzhaki, I. Klebanov, J. Maldacena,
S. Minwalla, L. Rastelli, K. Skenderis, A. Strominger,
N. Toumbas, A. Tseytlin, D. Vaman, H. Verlinde
and especially L. Motl for very 
useful discussions and comments. 
M.S. is supported by DOE grant DE-FG02-91ER40671, and
A.V. is supported by DE-FG02-91ER40654.

\appendix{A}{Some Matrices}

In this appendix we write down the matrices
$X^{(r)}$ which appear in the $\Delta$-functional
\Deltap.
We define $\beta = \alpha_{(1)}/\alpha_{(3)}$.  Then we consider
for $m,n>0$ the matrices of \GreenTC,
\eqn\adef{\eqalign{
A^{(1)}_{mn} &= (-1)^m \sqrt{n \over m} {2 \over \pi \alpha_{(1)}}
\int_0^{\pi \alpha_{(1)}}
d\sigma \cos {m \sigma \over \alpha_{(3)}} \cos {n \sigma \over
\alpha_{(1)}}\cr
&=(-1)^{m+n+1} {2 \sqrt{m n} \over \pi} 
{\beta \sin{m \pi \beta}\over n^2-m^2 \beta^2},
\cr
A^{(2)}_{mn} &= (-1)^m \sqrt{n \over m}
{2 \over \pi \alpha_{(2)}} 
\int_{\pi \alpha_{(1)}}^{\pi(\alpha_{(1)}+\alpha_{(2)})}
d\sigma\ \cos {m \sigma\over \alpha_{(3)}} \cos {n (\sigma - \pi \alpha_{(1)})
\over \alpha_{(2)}}
\cr
&=(-1)^{m+1} {2 \sqrt{m n} \over \pi} 
{(\beta+1) \sin{m \pi \beta} \over n^2-m^2 (\beta + 1)^2}
\cr
C_{mn} &= m \delta_{mn},
}}
and the vector
\eqn\bdef{
B_m = - {(-1)^m
\over \sqrt{m}} {2 \over \pi \alpha_{(1)} \alpha_{(2)}
} \int_0^{\pi \alpha_{(1)}}
d\sigma\ 
\cos {m \sigma \over \alpha_{(3)}}=
(-1)^{m+1} {2 \over \pi} {\alpha_{(3)} \over
\alpha_{(1)} \alpha_{(2)}} m^{-3/2} \sin m \pi \beta.
}
We define $X^{(3)}_{mn} = \delta_{mn}$, while
for $r=1,2$ we can express the matrix $X^{(r)}$ as
\eqn\aaa{
\eqalign{
X^{(r)}_{mn} &=
(C^{1/2} A^{(r)} C^{-1/2})_{mn}
\qquad\qquad\qquad{\rm if}~m,n>0,\cr
&= {\alpha_{(3)} \over \alpha_{(r)}} (C^{-1/2} A^{(r)} C^{1/2})_{
-m,-n},
\qquad{\rm if}~m,n<0,\cr
&=  - {1 \over \sqrt{2}} \epsilon^{rs} \alpha_{(s)} (C^{1/2} B)_m\qquad
\qquad~~~{\rm if}~n=0~{\rm and}~m>0,\cr
&=1\qquad\qquad\qquad\qquad\qquad\qquad\qquad{\rm if}~m=n=0,\cr
&= 0\qquad\qquad\qquad\qquad\qquad\qquad\qquad{\rm otherwise}.
}}
Schematically, the matrix $X_{mn}^{(r)}$ is given for $r=1,2$ by
\eqn\bigone{
{\myrm
\textfont1=\mymi
\textfont0=\myrm
\scriptfont0=\mysevenrm
\scriptscriptfont0=\myfiverm
\scriptfont1=\mysevenmi
\scriptscriptfont1=\myfivemi
\pmatrix{
\pmatrix{& {\myrm \vdots} & \cr \ldots & {\alpha_{(3)} \over \alpha_{(r)}}
(C^{-1/2}
A^{(r)}C^{1/2})_{-m,-n} & \ldots\cr & {\myrm \vdots} & } & & \cr
& & \cr
 & 1 & \cr
& & \cr
 & \pmatrix{ {\myrm \vdots} \cr  -{1 \over \sqrt{2}}
\epsilon^{rs} \alpha_{(s)} (C^{1/2} B)
_m\cr {\myrm \vdots}}& \pmatrix{& {\myrm \vdots} & \cr \ldots &
(C^{1/2} A^{(r)} C^{-1/2})_{mn} & \ldots\cr &{\myrm \vdots} &}
}}.
}

\appendix{B}{Harmonic Oscillator Integrations}

In this appendix we evaluate the integral
\eqn\apint{
E_a=
\int\left(\prod_{r=1}^3 \prod_{n=-\infty}^\infty
d p_{n(r)} \psi(p_{n(r)})
\right) \Delta \left[ \sum_{r=1}^3 p_{(r)}(\sigma)\right]
}
using \Deltap\ and \wavfun.
Overall factors of $2$ and $\pi$ will be dropped.
Let us define the matrix
\eqn\aaa{
(C_{(r)})_{mn} = \omega_{n (r)} \delta_{mn} = \sqrt{n^2 + (\alpha_{(r)}
\mu)^2} \delta_{mn}.
}
These matrices become singular in the flat-space limit $\mu \to 0$.
The quantity in \apint\ in parentheses may
be written as
\eqn\apinttwo{
\prod_{r=1}^3 dp_{(r)}\ 
\left( \det C_{(r)}\right)^{-1/4}
\exp \left[   - p_{(r)}^{\rm T} C^{-1}_{(r)} p_{(r)}
+ 2 a_{(r)}^{\dagger {\rm T}} C^{-1/2}_{(r)} p_{(r)}
- \ha a_{(r)}^{\dagger {\rm T}} a_{(r)}^\dagger
\right],
}
where we have introduced an obvious vector notation in order to
suppress the $n$ index which is summed from $-\infty$ to $+\infty$.
Let us make the change of variable
\eqn\aaa{
C^{-1/2}_{(r)} p_{(r)} \to p_{(r)}.
}
Then \apinttwo\ becomes
\eqn\apintthree{
 \prod_{r=1}^3 dp_{(r)}\ 
\left( \det C_{(r)}\right)^{1/4} \exp \left[
- p_{(r)}^{\rm T} p_{(r)} + 2 a_{(r)}^{
\dagger {\rm T}} p_{(r)} - \ha
a_{(r)}^{\dagger {\rm T}} a_{(r)}^\dagger \right].
}
Meanwhile, the $\Delta$-functional \Deltap\ becomes
\eqn\apdelta{
\delta\left( \sum_{r=1}^3 X^{(r)} p_{(r)}\right)
\to
\delta\left( \sum_{r=1}^3 X^{(r)} C_{(r)}^{1/2} p_{(r)}\right)
= \left( \det C_{(3)}\right)^{-1/2}  \delta \left(
\sum_{r=1}^3 \widetilde{X}^{(r)} p_{(r)}
\right),
}
where
\eqn\txdef{
\widetilde{X}^{(r)} = C_{(3)}^{-1/2} X^{(r)} C_{(r)}^{1/2}.
}
Using \apdelta\ to eliminate $p_{(3)}$ in
the integral \apintthree\ leaves
an integral of the form
\eqn\integral{
E_a= \mu_0 \int dp_{(1)} dp_{(2)}\ 
\exp\left[-\left(\matrix{ p^{\rm T}_{(1)} & p^{\rm T}_{(2)} }\right) M
 \left(\matrix{ p_{(1)} \cr p_{(2)} }\right)
+2 H^{\rm T}
\left(\matrix{ p_{(1)} \cr p_{(2)} }\right)
- \ha\sum_{r=1}^3 a^{\dagger {\rm T}}_{(r)} a^\dagger_{(r)}
\right],
}
where we have introduced
the determinant
\eqn\aaa{
\mu_0 = \left( {\det C_{(1)} \det C_{(2)}
\over \det C_{(3)}}\right)^{1/4},
}
the matrix
\eqn\aaa{
M = \left(\matrix{ 1 + \tX^{(1) {\rm T}} \tX^{(1)} &
\tX^{(1) {\rm T}} \tX^{(2)} \cr
\tX^{(2) {\rm T}} \tX^{(1)} & 1 + \tX^{(2) {\rm T}} \tX^{(2)}} \right)
}
and the vector
\eqn\hdef{
H = \left(\matrix{ a^\dagger_{(1)} - \tX^{(1) {\rm T}}a^\dagger_{(3)}
\cr
a^\dagger_{(2)} - \tX^{(2) {\rm T}} a_{(3)}^\dagger}\right).
}
Performing the Gaussian integral in \integral\ gives
\eqn\apintfour{
E_a = \mu_0 (\det M)^{-1/2} \exp\left[ 
H^{\rm T} M^{-1} H- \ha \sum_{r=1}^3 a^{\dagger {\rm T}}_{(r)}
a^{\dagger}_{(r)}\right].
}
Fortunately it is not to difficult to invert the matrix $M$.  We have
\eqn\minverse{
M^{-1} = \left(\matrix{ 1 - \tX^{(1) {\rm T}}
\widetilde{\Gamma}_a^{-1} \tX^{(1)} &
- \tX^{(1) {\rm T}} \widetilde{\Gamma}_a^{-1}\tX^{(2)} \cr
-\tX^{(2) {\rm T}} \widetilde{\Gamma}_a^{-1} \tX^{(1)} &
1 - \tX^{(2) {\rm T}} \widetilde{\Gamma}_a^{-1} \tX^{(2)}} \right)
}
in terms of the matrix
\eqn\aaa{
\widetilde{\Gamma}_a = \sum_{r=1}^3 \tX^{(r) } \tX^{(r) {\rm T}}.
}
Note that $\det \widetilde{\Gamma}_a = \det M$.
Combining \hdef, \apintfour\ and \minverse\ gives the final result
\eqn\earesult{
E_a = \mu_0 (\det M)^{-1/2}
 \exp \left[ {1 \over 2} \sum_{r,s=1}^3 a_{(r)}^{\dagger
{\rm T}} \overline{N}^{(rs)} a_{(s)}^\dagger \right],
}
where
the Neumann matrices are given by
\eqn\nmatrix{
\overline{N}^{(rs)}
= \delta^{rs} {\bf 1} - 2 \tX^{(r) {\rm T}} \widetilde{\Gamma}_a^{-1}
\tX^{(s)}.
}

One might be concerned that
\txdef\ singles out string number 3, since the vertex
should be symmetric under interchange of the three strings.
However, if we plug \txdef\ into 
\nmatrix, we find
that the factor of $C_{(3)}^{-1/2}$ drops out, leaving
\eqn\aaa{
\overline{N}^{(rs)} = \delta^{rs} {\rm 1} - 2 C_{(r)}^{1/2} X^{(r) {\rm T}}
{\Gamma}^{-1}_a
X^{(s)} C_{(s)}^{1/2}, \qquad
{\Gamma}_a = \sum_{r=1}^3 X^{(r)} C_{(r)} X^{(r) {\rm T}}.
}
Then the determinant which appears in \earesult\ may be written as
\eqn\aaa{
\mu_a =\mu_0 (\det M)^{-1/2} = \left( 
{ \det C_{(1)} \det C_{(2)}  \det C_{(3)}
}\right)^{1/4} (\det \Gamma_a)^{-1/2}.
}
Actually the vertex contains eight powers of $\mu_a$, one for
each transverse direction.

\listrefs

\end